\input harvmac
\input epsf

\newcount\figno
\figno=0
\def\fig#1#2#3{
\par\begingroup\parindent=0pt\leftskip=1cm\rightskip=1cm\parindent=0pt
\baselineskip=12pt
\global\advance\figno by 1
\midinsert
\epsfxsize=#3
\centerline{\epsfbox{#2}}
\vskip 14pt

{\bf Fig. \the\figno:} #1\par
\endinsert\endgroup\par
}
\def\figlabel#1{\xdef#1{\the\figno}}
\def\encadremath#1{\vbox{\hrule\hbox{\vrule\kern8pt\vbox{\kern8pt
\hbox{$\displaystyle #1$}\kern8pt}
\kern8pt\vrule}\hrule}}

\overfullrule=0pt

\noblackbox
\parskip=1.5mm

\def\Title#1#2{\rightline{#1}\ifx\answ\bigans\nopagenumbers\pageno0
\else\pageno1\vskip.5in\fi \centerline{\titlefont #2}\vskip .3in}

\font\caps=cmcsc10

\noblackbox
\parskip=1.5mm



           \def\CO{{\cal O}} 
\def\CA{{\cal A}}   
   
 \def\CR{{\cal R}}


\def\dj{\hbox{d\kern-0.347em \vrule width 0.3em height 1.252ex depth
-1.21ex \kern 0.051em}}

\def\half{{1\over 2}\,}

\def\ket{\rangle}
\def\bra{\langle}

\def\tX{\widetilde X}
\def\tphi{\widetilde \phi}
\def\tpi{\widetilde \pi}

\def\tPsi{\widetilde \Psi}
\def\tomega{\widetilde \omega}

\def\pt{\partial}

\def\Dirac{\,\raise.15ex\hbox{/}\mkern-13.5mu D}
\def\dirac{\,\raise.15ex\hbox{/}\kern-.57em \partial}
\def\aslash{\,\raise.15ex\hbox{/}\mkern-13.5mu A}
\def\btX{{\bf {\widetilde X}}}

\def\shalf{{\ifinner {\textstyle {1 \over 2}}\else {1 \over 2} \fi}}
\def\sshalf{{\ifinner {\scriptstyle {1 \over 2}}\else {1 \over 2} \fi}}
\def\sfourth{{\ifinner {\textstyle {1 \over 4}}\else {1 \over 4} \fi}}
\def\sthreehalfs{{\ifinner {\textstyle {3 \over 2}}\else {3 \over 2} \fi}}
\def\sdhalfs{{\ifinner {\textstyle {d \over 2}}\else {d \over 2} \fi}}
\def\sdmtwohalfs{{\ifinner {\textstyle {d-2 \over 2}}\else {d-2 \over 2} \fi}}
\def\sdmasonehalfs{{\ifinner {\textstyle {d+1 \over 2}}\else {d+1 \over 2} \fi}}
\def\sdmasthreehalfs{{\ifinner {\textstyle {d+3 \over 2}}\else {d+3 \over 2} \fi}}
\def\sdmastwohalfs{{\ifinner {\textstyle {d+2 \over 2}}\else {d+2 \over 2} \fi}}


\lref\comp{
L.~Susskind, L.~Thorlacius and J.~Uglum,
  ``The Stretched horizon and black hole complementarity,''
  Phys.\ Rev.\ D {\bf 48}, 3743 (1993)
  [hep-th/9306069].
  }
  
  \lref\fire{
   A.~Almheiri, D.~Marolf, J.~Polchinski and J.~Sully,
  ``Black Holes: Complementarity or Firewalls?,''
  JHEP {\bf 1302}, 062 (2013)
  [arXiv:1207.3123 [hep-th]].
  }
  
  \lref\usd{
  J.~L.~F.~Barbon and E.~Rabinovici,
  ``AdS Crunches, CFT Falls And Cosmological Complementarity,''
  JHEP {\bf 1104}, 044 (2011)
  [arXiv:1102.3015 [hep-th]].
  }
  
\lref\banksdo{
T.~Banks, M.~R.~Douglas, G.~T.~Horowitz and E.~J.~Martinec,
  ``AdS dynamics from conformal field theory,''
  hep-th/9808016.
  }
  
  \lref\susfrei{
   B.~Freivogel and L.~Susskind,
  ``A Framework for the landscape,''
  Phys.\ Rev.\ D {\bf 70}, 126007 (2004)
  [hep-th/0408133].
	  }
	  
\lref\susspull{
L.~Susskind,
  ``The Transfer of Entanglement: The Case for Firewalls,''
  arXiv:1210.2098 [hep-th].
  }
  
 \lref\eternalmalda{
 J.~M.~Maldacena,
  ``Eternal black holes in anti-de Sitter,''
  JHEP {\bf 0304}, 021 (2003)
  [hep-th/0106112].
  }

  \lref\mvr{
   M.~Van Raamsdonk,
  ``Building up spacetime with quantum entanglement,''
  Gen.\ Rel.\ Grav.\  {\bf 42}, 2323 (2010)
  [Int.\ J.\ Mod.\ Phys.\ D {\bf 19}, 2429 (2010)]
  [arXiv:1005.3035 [hep-th]].
  
  M.~Van Raamsdonk,
  ``A patchwork description of dual spacetimes in AdS/CFT,''
  Class.\ Quant.\ Grav.\  {\bf 28}, 065002 (2011).
  }
  
  \lref\mvrr{
  B.~Czech, J.~L.~Karczmarek, F.~Nogueira and M.~Van Raamsdonk,
  ``The Gravity Dual of a Density Matrix,''
  Class.\ Quant.\ Grav.\  {\bf 29}, 155009 (2012)
  [arXiv:1204.1330 [hep-th]].
  
   B.~Czech, J.~L.~Karczmarek, F.~Nogueira and M.~Van Raamsdonk,
  ``Rindler Quantum Gravity,''
  Class.\ Quant.\ Grav.\  {\bf 29}, 235025 (2012)
  [arXiv:1206.1323 [hep-th]].
  }

\lref\cdl{
 S.~R.~Coleman and F.~De Luccia,
  ``Gravitational Effects On And Of Vacuum Decay,''
  Phys.\ Rev.\  D {\bf 21}, 3305 (1980).

}

\lref\usu{
  J.~L.~F.~Barb\'on and E.~Rabinovici,
  ``Holography of AdS vacuum bubbles,''
  JHEP {\bf 1004}, 123 (2010)
  [arXiv:1003.4966 [hep-th]].}

\lref\mald{
  J.~Maldacena,
  ``Vacuum decay into Anti de Sitter space,''
  arXiv:1012.0274 [hep-th].
}

 \lref\adscft{
  J.~M.~Maldacena,
  ``The large N limit of superconformal field theories and supergravity,''
  Adv.\ Theor.\ Math.\ Phys.\  {\bf 2}, 231 (1998)
  [Int.\ J.\ Theor.\ Phys.\  {\bf 38}, 1113 (1999)]
  [arXiv:hep-th/9711200].
 S.~S.~Gubser, I.~R.~Klebanov and A.~M.~Polyakov,
  ``Gauge theory correlators from non-critical string theory,''
  Phys.\ Lett.\  B {\bf 428}, 105 (1998)
  [arXiv:hep-th/9802109].
 E.~Witten,
  ``Anti-de Sitter space and holography,''
  Adv.\ Theor.\ Math.\ Phys.\  {\bf 2}, 253 (1998)
  [arXiv:hep-th/9802150].
  }
  
  \lref\frag{
  
   J.~M.~Maldacena, J.~Michelson and A.~Strominger,
  ``Anti-de Sitter fragmentation,''
  JHEP {\bf 9902}, 011 (1999)
  [hep-th/9812073].
  }
  
  \lref\sen{
   A.~Sen,
  ``State Operator Correspondence and Entanglement in $AdS_2/CFT_1$,''
  Entropy {\bf 13}, 1305 (2011)
  [arXiv:1101.4254 [hep-th]].
  }

\lref\aff{
V.~de Alfaro, S.~Fubini and G.~Furlan,
  ``Conformal Invariance in Quantum Mechanics,''
  Nuovo Cim.\ A {\bf 34}, 569 (1976).
  }

\lref\kallosh{
P.~Claus, M.~Derix, R.~Kallosh, J.~Kumar, P.~K.~Townsend and A.~Van Proeyen,
  ``Black holes and superconformal mechanics,''
  Phys.\ Rev.\ Lett.\  {\bf 81}, 4553 (1998)
  [hep-th/9804177].
  }
\lref\mogollon{
S.~Hawking, J.~M.~Maldacena and A.~Strominger,
  ``DeSitter entropy, quantum entanglement and AdS/CFT,''
  JHEP {\bf 0105}, 001 (2001)
  [arXiv:hep-th/0002145].

A.~Buchel and A.~A.~Tseytlin, 
   ``Curved space resolution of singularity of fractional D3-branes on
  conifold,''
  Phys.\ Rev.\  D {\bf 65}, 085019 (2002)
  [arXiv:hep-th/0111017].

  A.~Buchel, P.~Langfelder and J.~Walcher,
  ``Does the tachyon matter?,''
  Annals Phys.\  {\bf 302}, 78 (2002)
  [arXiv:hep-th/0207235].

  A.~Buchel,
  ``Gauge / gravity correspondence in accelerating universe,''
  Phys.\ Rev.\  D {\bf 65}, 125015 (2002)
  [arXiv:hep-th/0203041].

A.~Buchel, P.~Langfelder and J.~Walcher,
  ``On time-dependent backgrounds in supergravity and string theory,''
  Phys.\ Rev.\  D {\bf 67}, 024011 (2003)
  [arXiv:hep-th/0207214].
  T.~Hirayama,
  ``A holographic dual of CFT with flavor on de Sitter space,''
  JHEP {\bf 0606}, 013 (2006)
  [arXiv:hep-th/0602258].
M.~Alishahiha, A.~Karch, E.~Silverstein and D.~Tong,
  ``The dS/dS correspondence,''
  AIP Conf.\ Proc.\  {\bf 743}, 393 (2005)
  [arXiv:hep-th/0407125].

  A.~Buchel,
  ``Inflation on the resolved warped deformed conifold,''
  Phys.\ Rev.\  D {\bf 74}, 046009 (2006)
  [arXiv:hep-th/0601013].

  O.~Aharony, M.~Fabinger, G.~T.~Horowitz and E.~Silverstein,
  ``Clean time-dependent string backgrounds from bubble baths,''
  JHEP {\bf 0207}, 007 (2002)
  [arXiv:hep-th/0204158].

  V.~Balasubramanian and S.~F.~Ross,
  ``The dual of nothing,''
  Phys.\ Rev.\  D {\bf 66}, 086002 (2002)
  [arXiv:hep-th/0205290].

  S.~F.~Ross and G.~Titchener,
  ``Time-dependent spacetimes in AdS/CFT: Bubble and black hole,''
  JHEP {\bf 0502}, 021 (2005)
  [arXiv:hep-th/0411128].

  R.~G.~Cai,
  ``Constant curvature black hole and dual field theory,''
  Phys.\ Lett.\  B {\bf 544}, 176 (2002)
  [arXiv:hep-th/0206223].

  V.~Balasubramanian, K.~Larjo and J.~Simon,
  ``Much ado about nothing,''
  Class.\ Quant.\ Grav.\  {\bf 22}, 4149 (2005)
  [arXiv:hep-th/0502111].

  J.~He and M.~Rozali,
  ``On Bubbles of Nothing in AdS/CFT,''
  JHEP {\bf 0709}, 089 (2007)
  [arXiv:hep-th/0703220].

 J.~A.~Hutasoit, S.~P.~Kumar and J.~Rafferty,
   ``Real time response on dS$_3$: the Topological AdS Black Hole and the
  Bubble,''
  JHEP {\bf 0904}, 063 (2009)
  [arXiv:0902.1658 [hep-th]].

}
\lref\dsconf{
  D.~Marolf, M.~Rangamani and M.~Van Raamsdonk,
  ``Holographic models of de Sitter QFTs,''
  arXiv:1007.3996 [hep-th].}

\lref\her{
  T.~Hertog and G.~T.~Horowitz,
  ``Holographic description of AdS cosmologies,''
  JHEP {\bf 0504}, 005 (2005)
  [arXiv:hep-th/0503071].
   T.~Hertog and G.~T.~Horowitz,
  ``Towards a big crunch dual,''
  JHEP {\bf 0407}, 073 (2004)
  [arXiv:hep-th/0406134].
}
\lref\insightfull{
  G.~Horowitz, A.~Lawrence and E.~Silverstein,
  ``Insightful D-branes,''
  JHEP {\bf 0907}, 057 (2009)
  [arXiv:0904.3922 [hep-th]].}

\lref\hartleh{
  J.~B.~Hartle and S.~W.~Hawking,
  ``Wave Function Of The Universe,''
  Phys.\ Rev.\  D {\bf 28}, 2960 (1983).
}

\lref\wbn{
  E.~Witten,
  ``Instability Of The Kaluza-Klein Vacuum,''
  Nucl.\ Phys.\  B {\bf 195}, 481 (1982).}

\lref\harsus{
  D.~Harlow and L.~Susskind,
  ``Crunches, Hats, and a Conjecture,''
  arXiv:1012.5302 [hep-th].}
\lref\har{
  D.~Harlow,
  ``Metastability in Anti de Sitter Space,''
  arXiv:1003.5909 [hep-th].}

\lref\eva{
 X.~Dong, B.~Horn, E.~Silverstein and G.~Torroba,
  ``Unitarity bounds and RG flows in time dependent quantum field theory,''
  Phys.\ Rev.\ D {\bf 86}, 025013 (2012)
  [arXiv:1203.1680 [hep-th]].
  
  X.~Dong, B.~Horn, E.~Silverstein and G.~Torroba,
  ``Perturbative Critical Behavior from Spacetime Dependent Couplings,''
  Phys.\ Rev.\ D {\bf 86}, 105028 (2012)
  [arXiv:1207.6663 [hep-th]].
	  }

\lref\polch{
I.~Heemskerk, D.~Marolf, J.~Polchinski and J.~Sully,
  ``Bulk and Transhorizon Measurements in AdS/CFT,''
  JHEP {\bf 1210}, 165 (2012)
  [arXiv:1201.3664 [hep-th]].
  }
  
  \lref\lifsh{
   A.~Hamilton, D.~N.~Kabat, G.~Lifschytz and D.~A.~Lowe,
  ``Local bulk operators in AdS/CFT: A Boundary view of horizons and locality,''
  Phys.\ Rev.\ D {\bf 73}, 086003 (2006)
  [hep-th/0506118].
  
  A.~Hamilton, D.~N.~Kabat, G.~Lifschytz and D.~A.~Lowe,
  ``Holographic representation of local bulk operators,''
  Phys.\ Rev.\ D {\bf 74}, 066009 (2006)
  [hep-th/0606141].
  
   A.~Hamilton, D.~N.~Kabat, G.~Lifschytz and D.~A.~Lowe,
  ``Local bulk operators in AdS/CFT: A Holographic description of the black hole interior,''
  Phys.\ Rev.\ D {\bf 75}, 106001 (2007)
  [Erratum-ibid.\ D {\bf 75}, 129902 (2007)]
  [hep-th/0612053].
  
  D.~Kabat, G.~Lifschytz and D.~A.~Lowe,
  ``Constructing local bulk observables in interacting AdS/CFT,''
  Phys.\ Rev.\ D {\bf 83}, 106009 (2011)
  [arXiv:1102.2910 [hep-th]].
  
}

\lref\bena{
 I.~Bena,
  ``On the construction of local fields in the bulk of AdS(5) and other spaces,''
  Phys.\ Rev.\ D {\bf 62}, 066007 (2000)
  [hep-th/9905186].
  
}

\lref\papado{
 K.~Papadodimas and S.~Raju,
  ``An Infalling Observer in AdS/CFT,''
  arXiv:1211.6767 [hep-th].
  }

\lref\BF{
  P.~Breitenlohner and D.~Z.~Freedman,
  ``Stability In Gauged Extended Supergravity,''
  Annals Phys.\  {\bf 144}, 249 (1982).

  P.~Breitenlohner and D.~Z.~Freedman,
  ``Positive Energy In Anti-De Sitter Backgrounds And Gauged Extended
  Supergravity,''
  Phys.\ Lett.\  B {\bf 115}, 197 (1982).
}
  
\lref\rmagan{
  J.~L.~F.~Barb\'on and J.~Mart\'{\i}nez-Mag\'an,
  ``Spontaneous fragmentation of topological black holes,''
  JHEP {\bf 1008}, 031 (2010)
  [arXiv:1005.4439 [hep-th]].
}

\lref\seiwit{
N.~Seiberg and E.~Witten,
  ``The D1/D5 system and singular CFT,''
  JHEP {\bf 9904}, 017 (1999)
  [arXiv:hep-th/9903224].
}

\lref\craps{
A.~Bernamonti and B.~Craps,
  ``D-Brane Potentials from Multi-Trace Deformations in AdS/CFT,''
  JHEP {\bf 0908}, 112 (2009)
  [arXiv:0907.0889 [hep-th]].
}


\baselineskip=15pt

\line{\hfill IFT UAM/CSIC-2013-084}
\line{\hfill CERN-PH-TH-2013-189}

\vskip 0.7cm

\Title{\vbox{\baselineskip 12pt\hbox{}
 }}
{\vbox {\centerline{Conformal Complementarity Maps
 }
}}

\vskip 0.5cm

\centerline{$\quad$ {\caps
Jos\'e L.F. Barb\'on$^\dagger$
 and
Eliezer Rabinovici$^\star$
}}
\vskip0.7cm

\centerline{{\sl  $^\dagger$ Instituto de F\'{\i}sica Te\'orica IFT UAM/CSIC }}
\centerline{{\sl  C/Nicolas Cabrera 13. 
 UAM, Cantoblanco 28049. Madrid, Spain }}
\centerline{{\tt jose.barbon@uam.es}}

\vskip0.2cm

\centerline{{\sl $^\star$
Racah Institute of Physics, The Hebrew University }}
\centerline{{\sl Jerusalem 91904, Israel}}
\centerline{{\sl and}}
\centerline{{\sl Theory Group, Physics Department, CERN}}
\centerline{{\sl CH 1211, Geneva 23. Switzerland}} 
\centerline{{\tt eliezer@vms.huji.ac.il}}

\vskip0.7cm

\centerline{\bf ABSTRACT}

 \vskip 0.3cm

 \noindent

We study quantum cosmological models for certain classes of bang/crunch singularities, using the duality between expanding bubbles in AdS with a FRW interior cosmology and perturbed CFTs on de Sitter space-time.  It is pointed out that horizon complementarity in the AdS bulk geometries is realized as a conformal transformation in the dual deformed CFT. The quantum version of this map is described  in full detail in a toy model involving conformal quantum mechanics. In this system the complementarity map acts as an exact duality between eternal and apocalyptic Hamiltonian evolutions. We calculate the commutation relation between the Hamiltonians corresponding to the different frames.
It vanishes only on scale invariant states.

\vskip 1cm

\Date{August 2013}

\vfill

\vskip 0.1cm




\baselineskip=15pt

\newsec{Introduction}

\noindent

There are quite a few cases in quantum field theory where the same set of data can be described in several ways. The two dimensional Ising model and electric magnetic duality are examples. In string theory a very large set of such relations was uncovered over the years.
Backgrounds which have for example different metric, topology, number of small or large dimensions and singularity, commutativity and associativity structures were identified. The AdS/CFT type relations are in this class.  In an effort to come to grips with the special challenges presented by black hole physics a concept named Black Hole Complementarity was put forward \refs\comp. Sets of observables defined outside and inside the horizon,  while not commuting among the sets, were each supposed to give a full description of the system.
The consequences of such a suggestion are still being processed (cf. \refs\fire\ and its wake). 

In a previous paper we have been brought to suggest a relation which touches all these types of dualities \refs\usd.
The systems discussed, a certain type of crunching AdS space-times, have a cosmological horizon separating those observers who meet the crunch in finite proper time from those who get to live for an infinite proper time.  The situation is thus similar to a black hole of infinite entropy. It was claimed  in \refs\usd\ that the exterior physics can be described, via  AdS/CFT  tools, by a specific class of non-singular time-independent QFTs living on a time-dependent de Sitter (dS) world volume, whereas the horizon interior could be described by a time-dependent QFT living on a static Einstein universe. The two holographic descriptions are related by a conformal transformation, which becomes equivalent to a complementarity map for this system. 

The conformal complementarity relates the `eternal' Hamiltonian evolution of dS space-time to a finite-time interval of the Einstein universe, which we call `apocalyptic'. This property is visible in the short-distance description of the QFT, and can be studied with effective Lagrangian methods, something we address in sections 2 and  3. Furthermore, if the system is simplified by `dimensional reduction' to the conformal quantum mechanics of de Alfaro, Fubini and Furlan \refs\aff, the complementarity transformation becomes explicitly  expressible for any wave function in the Hilbert space.   We analyze in sections 4 and 5 the details of this $d=1$ system, including the mapping of observables on both sides of the duality. Section 6 is devoted to a formal extension of the eternal/apocalyptic duality to arbitrary QFTs and we end with a discussion of conceptual puzzles and open questions in section 7, where we also succumb to the temptation to relate these ideas to some features of our universe.

\newsec{A Simple Model Of Cosmological Complementarity}

\noindent

A relation between horizon complementarity and conformal symmetry is inherent in AdS/CFT as a result of
basic rules of the correspondence. An AdS$_{d+1}$ space-time does not define a canonical metric on the $d$-dimensional boundary but rather defines  a boundary conformal structure, i.e. a conformal class of $d$-metrics. Conformal maps between these metrics extend naturally as bulk  diffeomorphisms, whose global  properties  produce  some degree of ambiguity in the precise rules by which a given abstractly defined  CFT codifies the bulk geometry.  

To appreciate the point, let us consider the AdS$_{d+1}$ global manifold with metric
\eqn\eframe{
ds_{\rm global}^2 = -(1+r^2)\, dt^2 + {dr^2 \over 1+r^2} + r^2 \,d\Omega_{d-1}^2\;,
}
where lengths are measured in units of the AdS radius of curvature. 
According to the standard AdS/CFT rules 
\refs\adscft, we may regard \eframe\ as the vacuum state of a dual CFT on the Einstein manifold ${\rm E}_d={
\bf R} 
\times {
\bf S}^{d-1}$ with metric 
\eqn\emet{
ds^2_{{\rm E}} = -dt^2 + d\Omega_{d-1}^2\;.
}
Small perturbations of \eframe\ quantized on a low-energy effective field theory approximation can be regarded as low-lying excitations of the CFT on \emet\ and can be described by  a Hamiltonian picture for all values of the Einstein-frame time variable $t
\in {
\bf R}$.

Alternatively, we could have started with a different presentation of the  AdS$_{d+1}$ space-time, with a metric which we denote `the bubble':
\eqn\dsframe{
ds_{\rm bubble}^2 = d\rho^2 + \sinh^2 (\rho) \left(-d\tau^2 + \cosh^2( \tau) \,d\Omega^2_{d-1} \right),}
made out of a de Sitter foliation of AdS. 
Taking the $\rho \rightarrow \infty $ limit and rescaling by the divergent factor of $\sinh^2 (\rho)$ we have a different conformal boundary metric: 
\eqn\dsm{
ds^2_{\rm dS} = -d\tau^2 + \cosh^2 (\tau)\,d\Omega_{d-1}^2
\;,}
given by the global de Sitter manifold. Thus, we can also regard the version of AdS given by \dsframe\ as the bulk dual of the CFT on the dS$_d$ global manifold with metric \dsm\ (cf. \refs\mogollon). Not surprisingly, the two boundary metrics are conformally related by a Weyl rescaling and a time diffeomorphism:
\eqn\conf{
ds^2_{{\rm dS}} = \Omega^2 (\tau) \,ds^2_{{\rm E}
} 
\;, \qquad \Omega (\tau) = \cosh (\tau) = {1\over \cos (t)}\;,}
a map that should be  unitarily represented in the Hilbert space and operator algebra of the abstractly defined CFT.

 \bigskip
\centerline{\epsfxsize=0.2\hsize\epsfbox{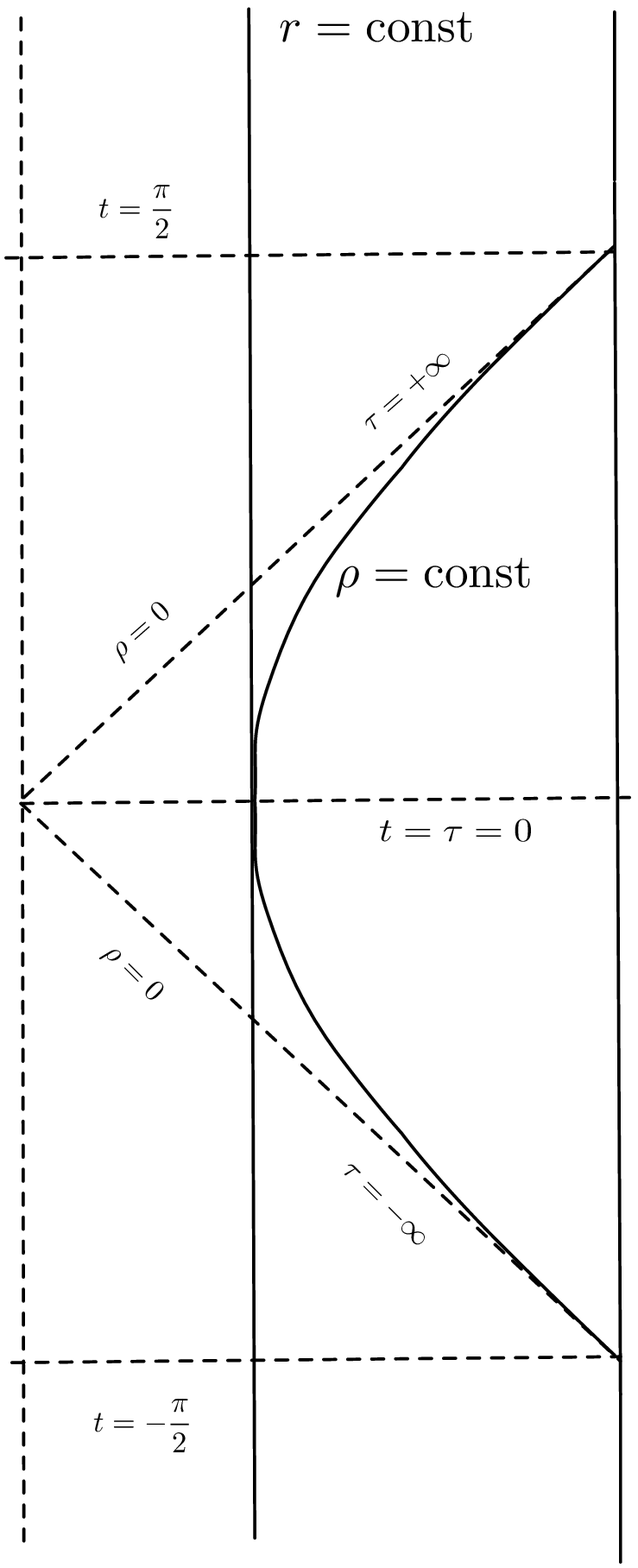}}
\noindent{\ninepoint\sl \baselineskip=2pt {\bf Figure 1:} {\ninerm
Radial slices adapted to E$_d$ and dS$_d$ isometries, corresponding to fixed $r$ and fixed $\rho$ respectively. Each point is a ${\bf S}^{d-1}$ sphere with radius ranging from zero at the origin of polar coordinates (left dashed line) to infinite size at the AdS boundary (right boundary line). }}
\bigskip

On the other hand,  the `bubble' version of AdS given by \dsframe, with coordinate domains  $-\infty < \tau<\infty$ and $0\leq \rho < \infty$, only covers a proper subset of the whole global AdS manifold \eframe: while the $r$-slices generate the whole AdS bulk, the $\rho$ slices only cover the causal diamond subtended by the $t\in  [-{\pi \over 2}, {\pi \over 2}]$ interval of E-time and bounded by the null surfaces $\rho=0$, $\tau = \pm \infty$.
 This raises the question of how the two CFT descriptions can be unitarily equivalent while one of the bulk duals is strictly contained into the other.

It turns out that the two bulk formulations are truly equivalent, in the sense that each one of them contains all the information needed to reconstruct the other \refs{\banksdo, \susfrei, \insightfull}. The key  fact making this equivalence possible is the existence of a common initial value surface in both bulk domains.
As shown in Figure 2, the Hamiltonian development of the dS-foliated patch shares an initial-value surface with the
E-foliation of the global AdS manifold, namely the $\tau=t=0$ surface. Therefore, any {\it perturbative} bulk state  defined on an arbitrary $t={\rm constant}$ surface may be unitarily  `pulled-back' to the $t=0$ initial-value surface, which coincides with the $\tau=0$ initial surface of the dS time slices. This operation is performed with the evolution operator generated by the $t$-Hamiltonian, the generator of translations in the foliation by $t=$ constant hyper-surfaces,   ${\widetilde H} \sim i\pt_t$.  Once the state is defined at $\tau=0$,
we may `push-forward' this state to any $\tau={\rm constant}$ surface in the dS patch, acting with the $\tau$-Hamiltonian $H \sim i\pt_\tau$. 

\bigskip
\centerline{\epsfxsize=0.4\hsize\epsfbox{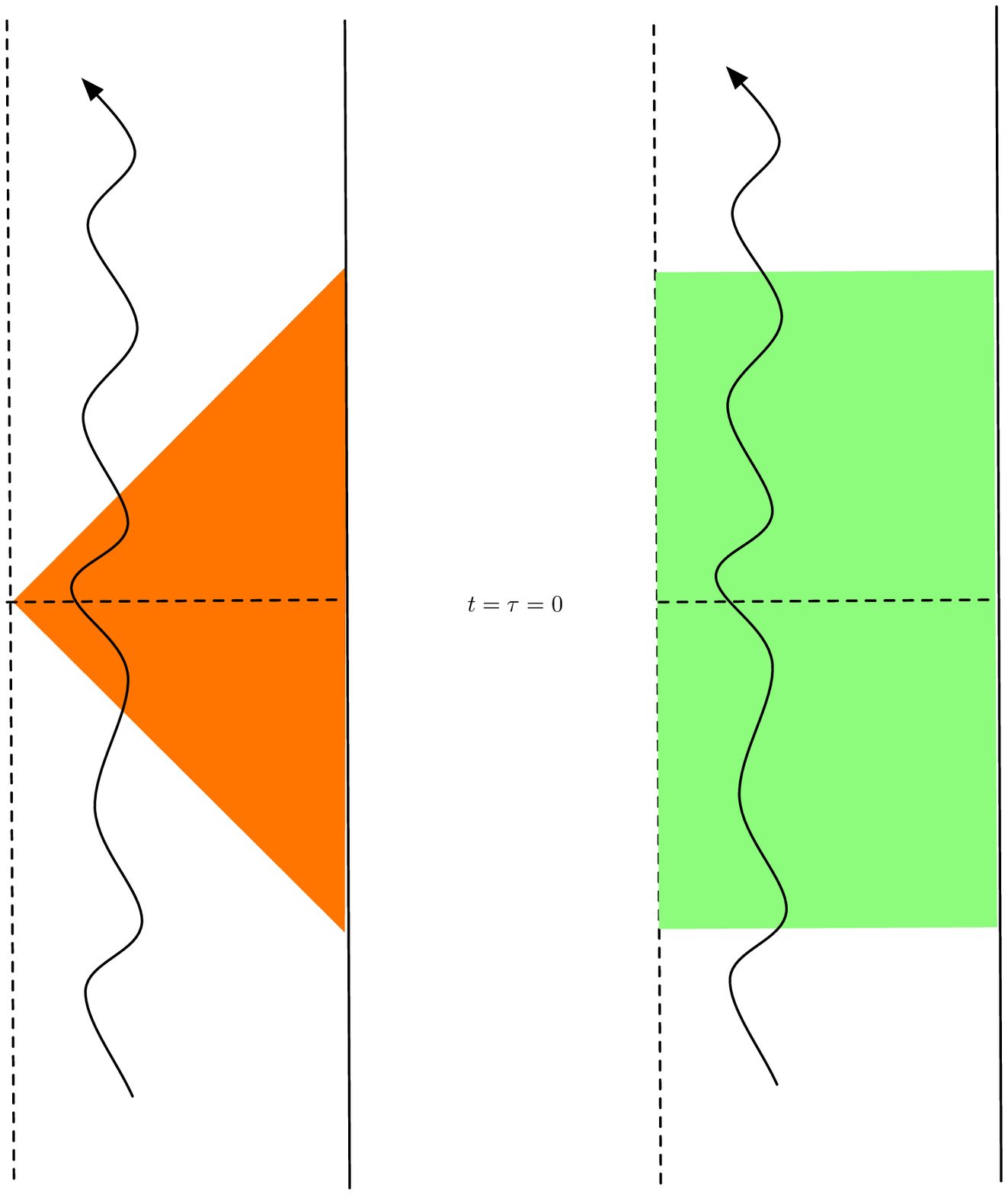}}
\noindent{\ninepoint\sl \baselineskip=2pt {\bf Figure 2:} {\ninerm
Domains of bulk AdS covered by the bulk Hamiltonian developments in dS time slicing (left) versus E time slicing (right), for the same interval of boundary data. Notice that both domains share the initial-value surface $t=\tau=0$. The wavy lines represent a perturbative particle-like state which can be propagated smoothly to all values of the $t$ variable. }}
\bigskip

\subsec{Extracting UV Data}

\noindent

The `pull-back/push-forward'  method described here (to follow the terminology of \refs\susspull), provides a simple operational definition of `horizon complementarity' in a very concrete  example. As it stands, the construction applies to  perturbative states around the vacuum AdS manifold. 

In seeking generalizations, it is natural to look at the asymptotic (UV) data, whose non-perturbative CFT interpretation is most straightforward. 
In this vein, we look for the effect on the AdS boundary of the alternative Hamiltonian foliations in the bulk and pick a natural map between $\tau={\rm constant}$ and $t={\rm constant}$ surfaces to represent the complementarity.  

\bigskip
\centerline{\epsfxsize=0.3\hsize\epsfbox{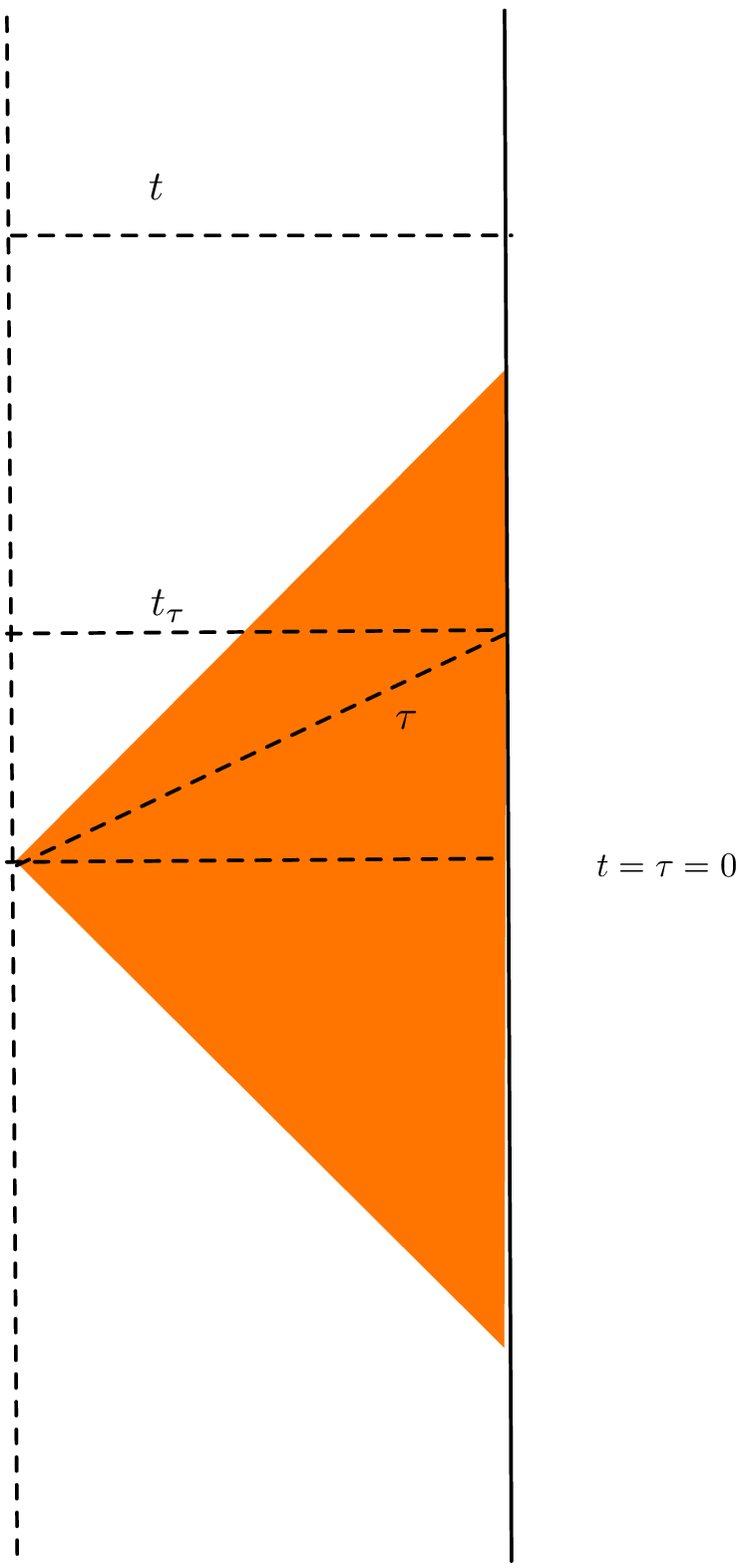}}
\noindent{\ninepoint\sl \baselineskip=2pt {\bf Figure 3:} {\ninerm
Any state specified by bulk data on fixed $t$ surfaces may be mapped unitarily into a state at fixed $\tau$ by pulling it back to $t=\tau=0$
as an intermediate step. The matching of time slices at the AdS boundary defines the time-diffeomorphism $t_\tau$. 
}}
\bigskip

  Directly matching the fixed-$t$ and fixed-$\tau$ surfaces at the AdS boundary (cf. Figure 3) provides such a natural map, determining a particular time-diffeomorphism which we shall denote $t_\tau$. 
We can find its explicit form using the common $SO(d)$ symmetry of \eframe\ and \dsframe\ to set $d\Omega_{d-1} =0$ in  both metrics.   Introducing coordinates $s= 
\tan^{-1} (r) - 
\pi/2$ and $u_\pm = 
\shalf (t \pm s )$ we obtain the metric of the AdS$_{2}$ section of \eframe: 
\eqn\muno{
ds_{1+1}^2 = -4{ du_+\, du_- \over \sin^2 (u_+ - u_-)}\;.
}
If instead we define $v_\pm = \shalf(\tau \pm \eta)$ with 
$$
\eta = \half \log\left({\cosh(\rho) -1 
\over \cosh(\rho) +1} \right) \;,
$$
we find a metric
\eqn\mudos{
ds_{1+1}^2 = -4{ dv_+ \,dv_- \over \sinh^2 (v_+ - v_-)}
}
for the $d\Omega_{d-1} =0$ section of \dsframe. By direct inspection, we can check that \muno\ and \mudos\ are related by the
transformation
\eqn\diftt{
\tan (u_\pm) = \tanh (v_\pm)
}
on their  domain of overlap. This includes 
 the AdS boundary, defined by $u_\pm = t/2$ and $v_\pm = \tau/2$. On this boundary, the  diffeomorphism \diftt\ reduces to the sought-for time-map:
\eqn\edst{
\tan(t_\tau/2) = \tanh(\tau/2)
\;,
} or, equivalently $
\cos (t_\tau) = 1/\cosh(\tau)$. The result is of course consistent with the conformal map between boundary dS$_d$ and E$_d$ metrics \conf.

Associating unitary evolution operators to the two time foliations we may write
\eqn\pp{
|\tPsi\ket_{t_\tau} =   {\widetilde U}_{t_\tau} \;U_\tau^{-1} \,|\Psi\ket_\tau\;
}
for the unitary map between the two states at fixed $t$ or fixed $\tau$ respectively. Our particular matching of time foliations, given by the diffeomorphism $t_\tau$, allows us to interpret \pp\ as the unitary implementation of the conformal map $\Omega$ between E$_d$ and dS$_d$, i.e. 
\eqn\uconf{
U_{\Omega} =    {\widetilde U}_{t_\tau} \,U_\tau^{-1} \;, \qquad \cos (t_\tau) = {1\over \cosh (\tau)}
\;.
}
Notice that $U_\Omega$ acts on the Hilbert space at a given value of the either time parameter\foot{ Alternatively, we may use the Heisenberg language, where the state is fixed conventionally as the $t=\tau=0$ state, and Hermitian operators representing observables are evolved unitarily in time. The two frames translate then into two operator algebras obtained from the action of the respective evolution operators on a given $t=\tau=0$ observable $A_0$. We have $
A(\tau) = U_\tau^{-1} \,A_0 \,U_\tau$, and ${\widetilde A}(t) = {\widetilde U}_t^{\;-1}\,A_0\,{\widetilde U}_t$. 
Again, the two operator algebras are unitarily related by the operator $U_\Omega ={\widetilde U}_{t_\tau} \, U_\tau^{\,-1}$.}, sending  a dS-frame state $|\Psi\ket_\tau$ at  dS time $\tau$ into an E-frame state $|\tPsi\ket_t$ with $t= t_\tau$. The singularity of this map at $t=\pm \pi/2$ does not translate into a physical singularity for perturbative states around the AdS vacuum. Those states  are perfectly smooth in the E-frame and may be continued for all values of $t\in {\bf R}$. The crucial issue of whether this smoothness is expected for more general states will be addressed in the next section.

The relation \pp\  was motivated by the geometry of the Hamiltonian flows in the AdS geometry,  and the evolution operators could be   constructed in  the low energy theory of the bulk,     describing perturbative states around the AdS vacuum manifold. However, the resulting operators are parametrized by time variables that make sense in the exact CFT, so it is natural to promote \pp\ as a non-perturbative definition of the `complementarity map'. 

\subsec{(In)Completeness}

\noindent

The same method can be implemented for the case of the Poincar\'e patch, 
\eqn\ppa{
ds^2_{\rm Poincare} = -y^2 \,dt'^{\,2} + {dy^2 \over y^2 } + y^2\,d\ell^2_{{\bf R}^{d-1}}\;,
}
defined for $t' \in {\bf R}$ and $y>0$.  The physics on this patch is codified by the Minkowski version of the CFT, i.e. picking a conformal boundary with metric
${\bf R}\times {\bf R}^{d-1}$, and the unitary complementarity map between \ppa\ and \eframe\ can be constructed as in \uconf, using the common $t'=t=0$ initial value surface.
 
An interesting example where this method {\it does not} work in a naive fashion is provided by the hyperbolic foliation of AdS:
\eqn\hy{
ds_{\rm hyp}^2 = -({\bar r}^{\,2} -1) \,d{\bar t}^{\,2} + {d{\bar r}^{\,2} \over {\bar r}^{\,2} -1} + {\bar r}^{\,2} \,d\ell^2_{{\bf H}^{d-1}}\;,
}
where the radial coordinate is defined in the domain ${\bar r} >1$ and the boundary metric is taken to be ${\bf R} \times {\bf H}^{d-1}$, the second factor being a $(d-1)$-dimensional hyperboloid. This time, the ${\bar t}=0$ surface does not cover the whole $t=0$ surface of the global manifold.  The null surface ${\bar r}=1$ is a horizon of a particular black hole solution with hyperbolic horizon geometry and Hawking temperature $T=1/2\pi$, which suggests that a situation similar to that of the eternal AdS black hole is at play
\refs\eternalmalda. Indeed, one can cover the complete initial value surface with the ${\bar t}=0$ section of  two hyperbolic patches of the form \hy, each one of them dual to the CFT living on  ${\bf R} \times {\bf H}^{d-1}$. 

The global AdS background is dual to the  CFT on the ${\bf S}^{d-1}$ vacuum, which can be regarded as an entangled state of the Hilbert spaces supported on each hemisphere of ${\bf S}^{d-1}$. Since a $(d-1)$-dimensional hemisphere is conformal to ${\bf H}^{d-1}$, let ${\cal U} $ denote the unitary operator implementing the map on the CFT Hilbert space. The global vacuum can be written then as an entangled state with data on  two copies of the hyperbolic CFTs (cf. \refs\mvr):
\eqn\vacent{
|{\rm VAC}_{{\bf S}^{d-1}} \ket = \sum_{E_{\rm hyp}} e^{-\pi E_{\rm hyp} } {\cal U}_L |  E_{\rm hyp} \ket_L \otimes {\cal U}_R | E_{\rm hyp} \ket_R\;,
}
where $E_{\rm hyp}$ is an energy eigenvalue of the CFT quantized on the ${\bf H}^{d-1}$ spatial manifold\foot{This correspondence contains subtleties for CFTs with scalars, which have negative mass-squared due to the conformal coupling to the negative scalar curvature of ${\bf H}^{d-1}$. When the theory is defined on a compact quotient of  ${\bf H}^{d-1}$ there are genuine tachyonic zero modes and non-perturbative instabilities (cf. \refs\rmagan)} (the sum in \vacent\ is symbolic, since the spectrum of hyperbolic energies is continuous). 
Operators on a single copy see the AdS vacuum as a mixed thermal state with the temperature
$T=1/2\pi$ of the hyperbolic AdS black hole. \foot{Notice that ${\bf R}\times {\bf H}^{d-1}$ is conformal to the static patch of dS$_d$. Hence we are consistent with the global E$_d$/dS$_d$ map, since two static dS patches are needed to cover the sphere ${\bf S}^{d-1}$ at $\tau=0$ in the global dS space. }

 \bigskip
\centerline{\epsfxsize=0.5\hsize\epsfbox{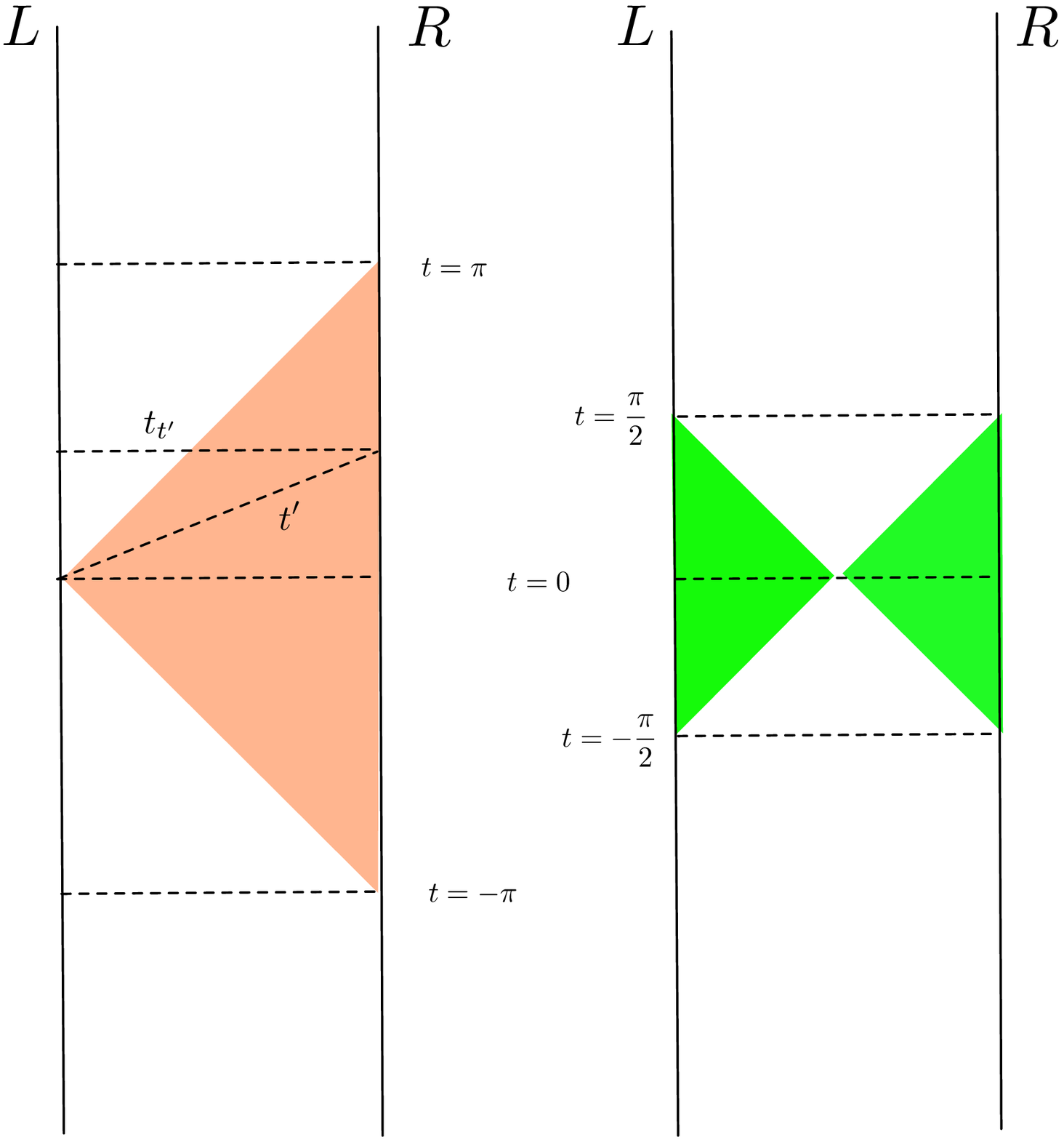}}
\noindent{\ninepoint\sl \baselineskip=2pt {\bf Figure 4:} {\ninerm
Causal diagram of the Poincar\'e patch (left figure) and the hyperbolic patches (right figure) in AdS. Unlike the previous global representations of AdS$_{d+1}$, the line denoted $L$ is a true boundary component, rather than the origin of polar coordinates and points represent surfaces homeomorphic to ${\bf R}^{d-1}$ rather than spheres. For $d>1$ the two boundary components, denoted $L$ and $R$ in the figure,  define  subsets  of the complete conformal boundary  ${\bf R} \times {\bf S}^{d-1}$. 
For $d=1$ this picture gives a complete representation of the causal structure, where the two boundary  components $L$ and $R$ are truly disconnected. 
}}
\bigskip

An important comment regarding \vacent\ is that, while the ${\bf S}^{d-1}$ vacuum state $|{\rm VAC}_{{\bf S}^{d-1}} \ket$ of the CFT  should map smoothly to the global AdS geometry, the same cannot be said of each individual eigenstate of the hyperbolic Hamiltonian $|E_{\rm hyp} \ket$. As emphasized in \refs\mvr, such states are expected to harbor bulk singularities (akin to `firewalls' \refs\fire) on the  horizon of the hyperbolic patch. In the next section we shall add a simple classical argument in favor of this interpretation.

Complementarity maps from a left-right symmetric slicing in hyperbolic time ${\bar t}$  to some global E-frame slice, $t$,  can be specified by operators of the form 
\eqn\aco{
U_{\rm C}  ={\widetilde {U}}_{t}  \left( {\cal U}_L^{-1} U_{
\bar t}^{-1}  \otimes  {\cal U}_R^{-1} U_{\bar t}^{-1}  \right)
\;.
}
In this expression, the first factor pulls the fixed-${\bar t}$ state in the product hyperbolic CFT  back into the ${\bar t}=0$  slices, undoes the conformal map back to each left-right hemispheres and finally it pushes  the full ${\bf S}^{d-1}$ state forward in E-frame time $t$.  Notice, however, that $U_{\rm C}$ defined in \aco\  makes use of the two copies of the CFT on disjoint hyperboloids, and the resulting operator does not have a straightforward interpretation as a unitary representation  of a conformal map in the full CFT defined on ${\bf R}\times {\bf S}^{d-1} $.

\subsec{$d=1$}

\noindent

We note that the  $d=1$ case has interesting peculiarities.    The union of the back-to-back hyperbolic patches of AdS$_{1+1}$  coincides
with the bubble patch. Their boundary is dS$_1$, consisting of two disconnected lines, each one representing one static dS patch (cf. Figure 4).  The map between Poincar\'e and global frames also simplifies. We compute here for future use  the associated boundary  time diffeomorphism. 
Let the Poincar\'e patch of AdS$_{1+1}$ be represented by the metric
\eqn\pin{
ds^2_{\rm Poincare} = -y^2 \,dt'^{\,2} + {dy^2 \over y^2}\;,
}
with $y\geq 0$ and $t' \in {\bf R}$, covering a proper subset of the global AdS$_{1+1}$ whose metric  we write as 
\eqn\glod{
ds^2_{\rm global} = -(1+x^2) \,dt^2 + {dx^2 \over 1+x^2} \;,
}
with $x \in {\bf R}$ and $t\in {\bf R}$, the right and left boundaries corresponding to the limits $x\rightarrow \pm \infty$ respectively.  As indicated  in Figure 4, a natural time-diffeomorphism $t_{t'}$ is induced on the boundary metrics by the matching of time slices at the R boundary $x=y=+\infty$. To find this boundary diffeomorphism we begin by transforming \pin\ by the change of variables 
$\zeta_\pm = t' \pm 1/y$, leading to
\eqn\pg{
ds^2 = -4\,{d\zeta_+ \,d\zeta_- \over (\zeta_+ - \zeta_-)^2}\;,
}
which in turn may be transformed into the global version \muno\ under the further redefinition $\zeta_\pm = \tan(v_\pm)$. Evaluating the chain of coordinate changes at the R boundary, we find 
\eqn\fc{
t' = \tan (t_{t'}/2)\;
}
for the required time-diffeomorphism. 

It is tempting to promote the picture of complementarity maps outlined in this section to conformal maps in CFT$_1$, i.e. a model of conformal quantum mechanics which would encode the physics of AdS$_{1+1}$ spaces. On the other hand, the $d=1$ version of the AdS/CFT correspondence is rich with subtleties (cf. for instance \refs{\frag, \sen}) which makes it a rather special case. Despite these caveats, we will find in the coming sections that many aspects of the complementarity maps discussed here do find analogs in the simplest models of conformal quantum mechanics.

\newsec{Singular Maps Versus Singular States}

\noindent 

We have argued that a version of horizon complementarity for perturbative bulk states around the global AdS vacuum can be analyzed in terms of conformal maps between the E$_d$ and dS$_d$ versions of the dual CFT. This conformal rescaling, which we refer to as the EdS map, sends the whole Hamiltonian development of the dS  manifold into a compact domain of Einstein-frame time. We refer to this situation as the `eternal/apocalyptic duality'. Accordingly, we speak of the `eternal Hamiltonian', dual to the dS time variable, $\tau$, and the `apocalyptic Hamiltonian', dual to the E-frame time variable, $t$. The conformal transformation $U_\Omega$ is singular at the endpoints of apocalyptic time $t=t_\star=\pm \pi/2$, but the physics of perturbative states around AdS is smooth, as the E-frame Hamiltonian acts smoothly on those states for $|t|> \pi/2$. 

It is possible to envisage states without such a  smooth continuation, for which the apocalyptic time development  is truly singular in a physical sense. Let us consider a classical state with the properties of a codimension-one brane,  supported on a fixed  
$\rho$ trajectory in \dsframe. Such a state is stationary with respect to the
$\tau$-Hamiltonian, but it is  accelerating, asymptotic to a null surface,   in the E-frame of the global AdS geometry. Therefore it requires an infinite supply of $t$-energy, and its $t$-time evolution is not expected to be smooth  for $\Delta t > \pi$. An example of this behavior
is given by a $O(d,1)$-invariant configuration similar to a  Coleman-de Luccia  (CdL) bubble, which expands exponentially in an ambient AdS space and produces a crunch as in Figure 5. \foot{These solutions can be constructed as Lorentzian continuations, in the sense of Hartle and Hawking \refs\hartleh, of $O(d+1)$-invariant Euclidean solutions with the interpretation of renormalization-group flows of the Euclidean CFT on  ${\bf S}^d$ (cf. \refs\usd).}

Brane-like states producing crunch singularities are a rather more interesting arena where ideas of complementarity can be    probed. Since the whole space-time crunches, they behave in some sense as infinite-entropy limits of black holes --even the boundary of AdS `crunches' in finite global time.  Local observables associated to constant-$\rho$ trajectories are analogous to `exterior' black hole observables, whereas local observables associated to constant-$r$ trajectories are analogous to `infalling' observables. Unlike the black hole case, we can identify infalling `observers' even on the AdS boundary, so that the complementarity map must be visible in the deep UV data of the CFT. 

 \bigskip
\centerline{\epsfxsize=0.2\hsize\epsfbox{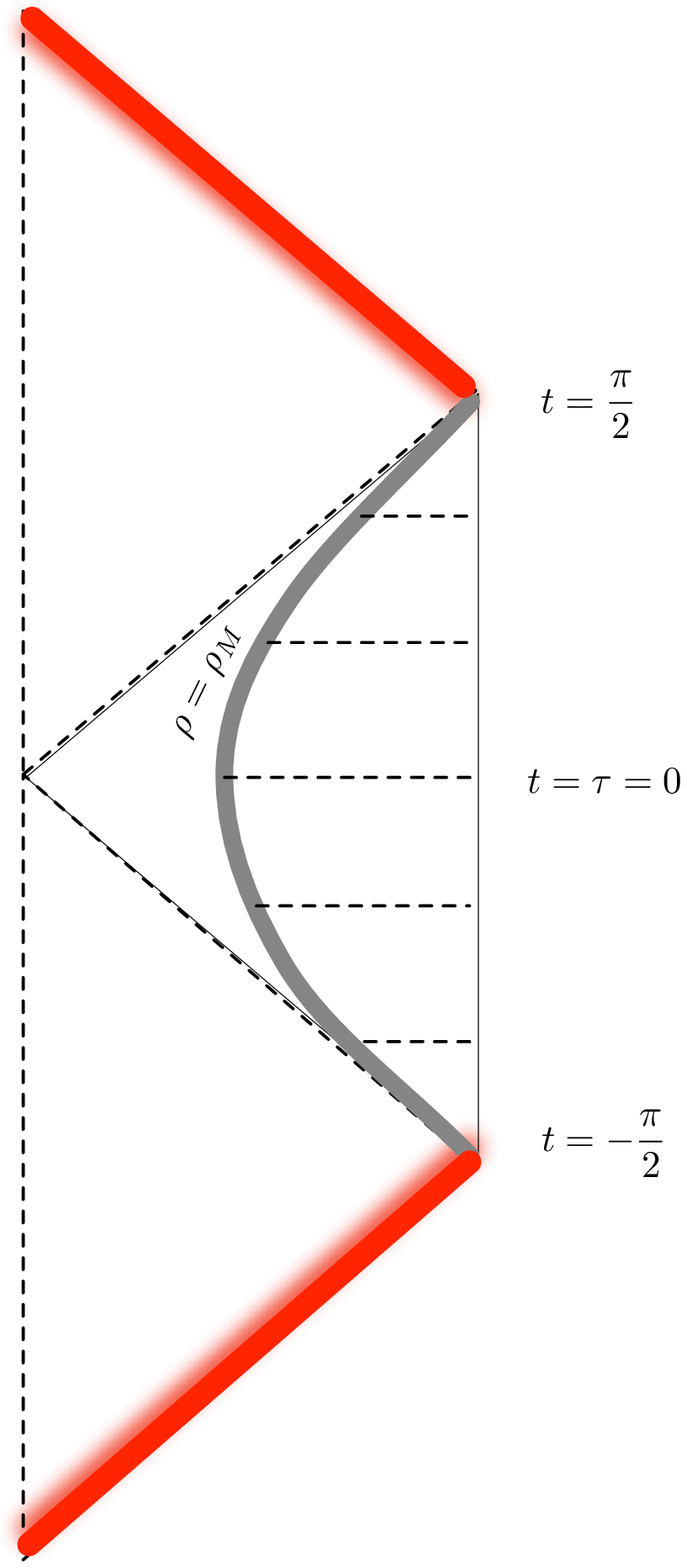}}
\noindent{\ninepoint\sl \baselineskip=2pt {\bf Figure 5:} {\ninerm
A  $O(d,1)$ invariant bubble of finite dS energy, producing a crunch at $t=\pm \pi/2$. Surfaces of fixed $t$ in the exterior AdS geometry are indicated in the picture. If the shell is very thin, the interior geometry is also well approximated by AdS except near the bang/crunch singularities.
}}
\bigskip

 It is precisely the conformal transformation between `eternal' and `apocalyptic' Hamiltonian flows  what provides this    `UV remnant' of the complementarity map, visible in the microscopic formulation of the CFT.  
In what follows, we study the transformation between eternal and apocalyptic Hamiltonians from various points of view, starting with a Landau--Ginzburg description of the codimension-one brane states.

\subsec{Effective Landau--Ginzburg Models}
\noindent

          An approximate  description of  $O(d,1)$-invariant brane states can be achieved by defining a radial collective coordinate $\phi$ which can be regarded as a field degree of freedom in the CFT. Assuming that this world-volume field is weakly coupled, it can be assigned a   canonical mass  dimension. A brane situated at $\rho=\rho_M$ can be expressed by arranging the effective dynamics such that the collective field
$\phi$ obtains an expectation value
\eqn\conds{
\bra \phi \ket_M \sim M^{d-2 \over 2}\;,
}
where $M$ is the mass scale associated to the fixed radial position $\rho=\rho_M$. According to the IR/UV relation of AdS/CFT 
we have  (cf. \refs\usd) 
\eqn\uvird{
\rho_M 
\sim 
\log \,\bra \phi\ket_M \sim \log (M)\;,
}
a relation which is valid provided $d >2$ and  $M\gg 1$ in units of the dS curvature radius, two conditions that we assume to be valid throughout this section.

The simplest effective dynamics supporting such a classical condensate on dS is given by the effective (long wavelength) Landau--Ginzburg (LG) action 
\eqn\effl{ 
S[\phi]_{\rm eff} = -\int_{{\rm dS}_d}\left[ \shalf |\pt \phi |^2 + V_{\rm eff} (\phi) \right] \;
,}
where the effective potential can be written as 
\eqn\modlg{
V_{\rm eff} [\phi] = 
 \shalf \xi_d\, {\cal R}_{{\rm dS}_d} \,\phi^2 + \lambda\,\phi^{2d \over d-2} + \varepsilon M^{d-\Delta} \phi^{2\Delta \over d-2} \;.
}
The  first term and the marginality of the operator appearing in the second term are dictated by conformal invariance\foot{These terms are induced at large $\rho_M$ from the Dirac--Born--Infeld action of the brane (cf. \refs{\seiwit, \usu})}, including the conformal curvature coupling with 
$$
\xi_d = {d-2 \over 4(d-1)}\;.
$$
The non-linear terms in \modlg\ correspond to a marginal operator of mass dimension $d$ and a relevant operator of dimension $\Delta < d$, whose coupling introduces the conformal symmetry-breaking  scale $M$. The factor $\varepsilon = \pm 1$ controls the sign of the relevant operator, and we must require $\lambda >0$ for global stability.  In general, there may be many relevant operators and a host of irrelevant operators correcting \effl, but the simplified form of \modlg\ will suffice for our  qualitative discussion.

Taking $\lambda = \CO(1)$ and $M\gg 1$ we can find condensates of the form \conds\ provided $\varepsilon =-1$. In fact, we get both a stable condensate and an unstable one, as shown in Figure 6. The unstable condensate was interpreted in \refs{\usu, \usd} as  a sphaleron configuration which all the properties of a CdL bounce in the bulk. Interestingly, this configuration is present even for the globally unstable model with no relevant operator,   $M=0$ and $\lambda<0$. Such models  were studied extensively in \refs{\her, \craps,  \mald, \har, \usu, \usd} as holographic duals of crunch singularities. It was recognized in \refs{\mald, \harsus,  \usd} that
the stable condensates in globally well-defined models are perfectly suited to  the AdS/CFT embedding of space-times with crunch singularities.

The classical LG description of condensate states on dS should be accurate when the scale of the condensate is much larger than the dS temperature, i.e. $M\gg 1$ in our notation. In the opposite limit, $M\ll1$, the effective LG theory should receive large quantum corrections. On the other hand, this is the limit where classical gravity descriptions in the bulk admit a linearized approximation (cf. the appendix of \refs\mald), the result being $O(d,1)$-invariant geometries with very small bubbles and the same crunching behavior as in Figure 4

 \bigskip
\centerline{\epsfxsize=0.4\hsize\epsfbox{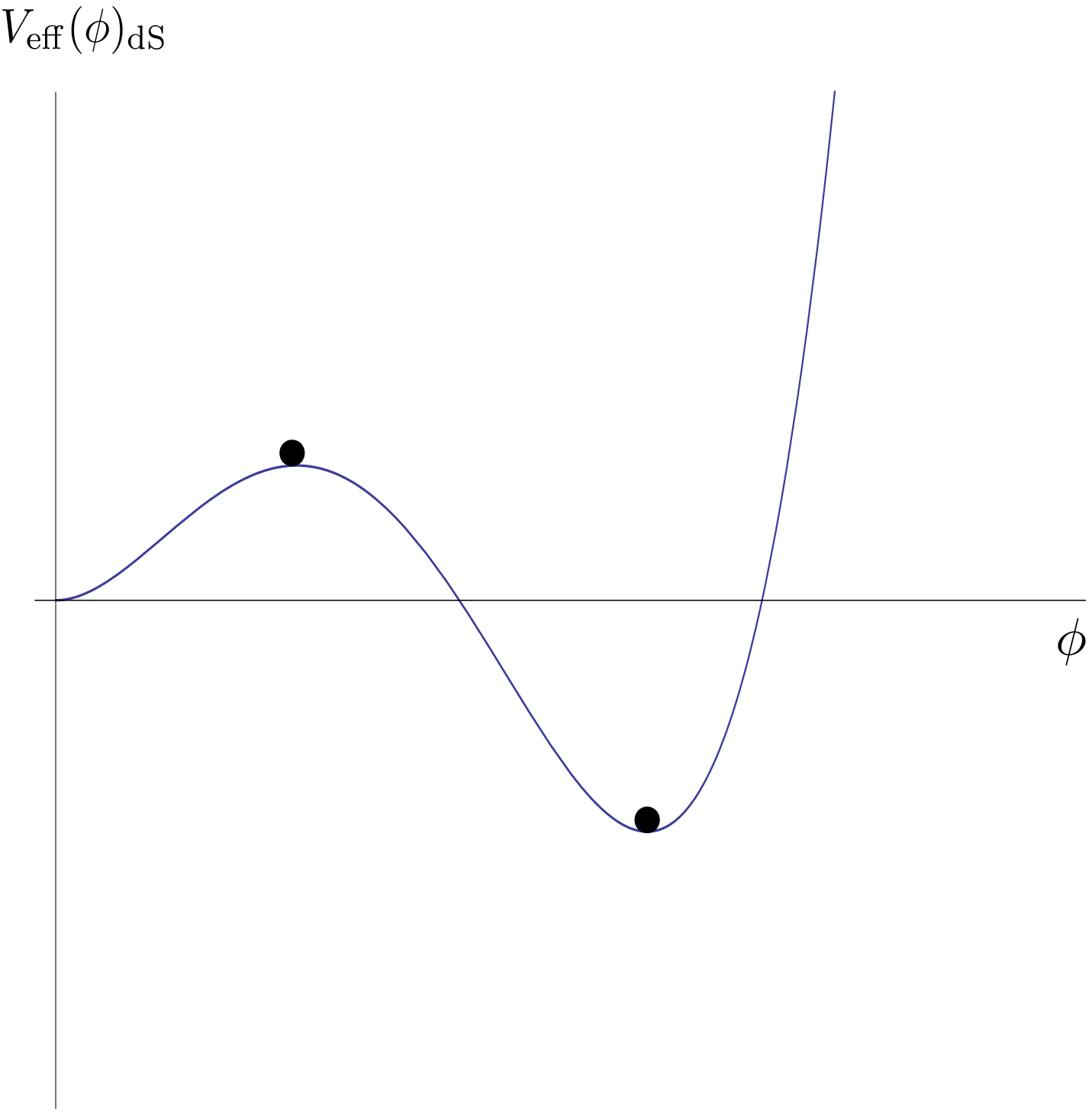}}
\noindent{\ninepoint\sl \baselineskip=8pt {\bf Figure  6:} {\ninerm
Schematic representation of the de Sitter effective LG potential with $M\gg 1$ and $\varepsilon =-1$. The unstable condensate at the local maximum is dual to a CdL bubble in the bulk. }}
\bigskip

The conformal complementarity (EdS) map  \uconf\ becomes particularly simple in the classical approximation to the LG dynamics \effl. Given the conformal map between the two frames
\conf, an extension to the full effective LG field dynamics is achieved by postulating the conformal transformation of  the basic field variable, as dictated by its scaling dimension: 
\eqn\mad{
\tphi(t)= \Omega(t)^{d-2 \over 2} \,\phi(\tau_t)\;.
}
This transformation sends the dS-invariant condensate $\bra \phi \ket_M \sim M^{d-2 \over 2}$ into the $t$-dependent E-frame configuration 
\eqn\madd{
\bra \tphi (t)\ket_M  \propto \left({M \over \cos (t)}\right)^{d-2 \over 2}\;,
}
which is a solution of the E-frame system
\eqn\efsys{
{\widetilde S}_{{\rm E}_d} = \int_{{\rm E}_d} \left( \shalf |\pt \tphi |^2 + \shalf \xi_d {\cal R}_{{\rm E}_d} \,\tphi^{\,2}+ {\widetilde V}_{\rm eff} (\tphi\,) + \dots\right) \;,
}
with an effective potential 
 \eqn\epof{
 {\widetilde V}_{\rm E} [\tphi\,] = \shalf \xi_d\,{\cal R}_{\rm E} \,\tphi^2 + \lambda \,\tphi^{\,{2d \over d-2}} -\left({M \over \cos(t)}\right)^{d-\Delta} \,\tphi^{\,{2\Delta \over d-2}}\;
 }
 featuring an explicit time-dependent coupling of the relevant operator.  This coupling causes the 
  total energy, as well as the kinetic and potential energies of the state $\tphi(t)$ to blow up at the `bang-crunch' times $t_\star = \pm \pi/2$, showing that the singularities at the `apocalyptic' times are physical in terms of the E-frame variables. The E-frame Hamiltonian is itself singular at $t=t_\star$, so that the $t$ time  evolution cannot possibly continue smoothly beyond the apocalyptic times. 

We conclude that a  particular class of $O(d,1)$-invariant  states in dS-frame variables  are seen as a singular (crunching) states in the E-frame, as a result of a singular driving term in the E-frame Hamiltonian. 

By inverting \mad\ we can study how an E-frame $t$-stationary  condensate looks when analyzed in dS-frame variables. Such states have $U(1)\times O(d)$ symmetry and have the form $\bra \tphi\,\ket_{\widetilde M}  \propto {\widetilde M}^{\,{d-2 \over 2}}$  (notice that we now take $\pt_t {\widetilde M} =0$). This  $t$-static configuration is a solution of the static E-frame potential
$$
{\widetilde V}_{\rm E} [\tphi] = \shalf \xi_d\,{\cal R}_{\rm E} \,\tphi^2 + \lambda \,\tphi^{\,{2d \over d-2}} - {\widetilde M}^{\,d-
\Delta} \,\tphi^{\,{2\Delta \over d-2}}\;.
$$
  The corresponding dS-frame field is
\eqn\maddd{
\phi_{\widetilde M} (\tau) \propto \left( {{\widetilde M} \over \cosh (\tau)}\right)^{d-2\over 2}\;, 
}
which vanishes exponentially in global dS time for $d>2$.  After appropriately normalizing the $\CO(1)$ proportionality constant in \maddd, this solution is driven by the dS-frame LG `potential' 
\eqn\dslg{
V_{\rm dS} [\phi] = \shalf \xi_d \,\CR_{{\rm dS}_d} \,\phi^{\,2} + \lambda \,\phi^{\,{2d\over d-2}} - \left({{\widetilde M} \over \cosh(\tau)}\right)^{d-\Delta} \,\phi^{\,{2\Delta \over d-2}}\;,
}
which now features a negative-definite, $\tau$-dependent relevant operator turning-off as $|\tau| \rightarrow \infty$. The value of
the LG potential evaluated on the solution \maddd\  also redshifts to zero as $|\tau|\rightarrow \infty$. 
We thus conclude
that the $U(1)\times O(d)$-invariant  condensates on the E-frame  `dilute away' when analyzed in dS-frame variables. 

Broadly speaking, we can identify two qualitatively different types of states.
One natural class is given by those states which are completely smooth in the E-frame and can be continued through all $t\in {\bf R} $ with a time-independent non-singular E-frame Hamiltonian. We refer to these as {\it smooth} states. When analyzed in the eternal frame, their distinctive feature is the `diluting' nature as $|\tau| \rightarrow \infty$. 

A second class of states is given by those which are asymptotically $\tau$-stationary in the eternal frame, but distinct from the trivial CFT vacuum on dS. The natural way of engineering such states is to deform the CFT by a relevant operator and consider stationary states looking like condensates induced by the new relevant interactions.
These states, while completely regular in the eternal dS frame, are singular in the E-frame and therefore called {\it crunch} states.  

It should be clear that the smooth and crunch states do not share the same phase (or Hilbert) space. They actually occur in different systems, in the sense that they need different Hamiltonians to be supported as stationary states. If we fix, say the dS frame, crunch states need a non-trivial dS-invariant relevant deformation to be turned on, while smooth states already exist in dS systems whose Hamiltonian has no such deformation turned on. 

We have chosen to discuss the conformal map which rises naturally from the diffeomorphisms discussed in section 2.
It maps a non compact region into a compact one independent of the dynamics brought about by the specific
Hamiltonian involved. This more universal approach required us to disentangle the singularity inherent in such a transformation from a possible dynamical one. One could have chosen a conformal transformation akin to a unitary
gauge in gauge theories.  It would be {\it ab initio} useful when there is a physical singularity to be exposed in one frame,
like the unitary gauge is useful in the BEH phase. The transformation will be defined on the fields
(cf.  equation \mad) in such a way that  $ \Omega$  is multiplied by  the product of the expectation value of
the scalar field $\phi$ in the dS frame and the Hubble scale. This product  vanishes in the cases when there is no condensate and thus renders the transformation to be ill defined in those cases.

\subsec{Classical Firewalls}

\noindent

It is interesting to inquire to what extent this  description generalizes to other versions of the conformal frame duality studied here, such as the examples of AdS foliations related to CFTs on flat or hyperbolic space-times.  

Let ${\bf K}_k$ represent the standard constant-curvature manifold in $d-1$ dimensions, with $k=0, \pm 1$ controlling the sign of the Ricci curvature, i.e. ${\bf K}_0 = {\bf R}^{d-1}$, ${\bf K}_1 = {\bf S}^{d-1}$ and ${\bf K}_{-1} = {\bf H}^{d-1}$.   We can describe the global,  Poincar\'e and hyperbolic patches of AdS at once with the family of metrics:  
 \eqn\manyfol{
ds^2_k = -(r_k^2 + k)\,dt_k^2 + {dr_k^2 \over r_k^2 + k} + r_k^2 \,d\ell^2_{{\bf K}_k}\;.
}
The  $k=1$ case with $r_1 \geq 0$ 
is the standard metric of the global AdS manifold \eframe. The case  $k=0$ with $r_0 \geq 0$  gives the Poincar\'e patch \ppa\ of AdS, and finally $k=-1$  with $|r_{-1}| \geq 1$ returns 
 the two mirror hyperbolic patches given by  \hy. The time variables $t_k$ in 
\manyfol\ define natural Hamiltonian flows for CFTs on ${\bf R} \times {\bf K}_k$. In the notation of the previous section, we
have $t=t_1$, $t' = t_0$ and ${\bar t}= t_{-1}$.

   It is interesting to inquire about the fate of condensate states with the symmetries of ${\bf R} \times {\bf K}_k$, corresponding to brane-like states defined by $r_k = {
   \rm constant} $ in \manyfol. In particular, one can consider condensates on ${\bf R}\times {\bf R}^{d-1}$ with Poincar\'e invariance $ISO(d-1, 1)$ and condensates on ${\bf R} \times {\bf H}^{d-1}$ with symmetry $U(1) \times O(d-2, 1)$. \foot{Interestingly, these condensates can be defined without the need of a relevant operator. The $k=0$ flat-space case is similar to a Higgssed state on a Coulomb branch and the $k=-1$ hyperbolic case only requires the existence a positive marginal deformation to stabilize the negative-definite conformal mass term.}

The crucial property making the $k=0$ and $k=-1$ cases special is the non-compact nature of the spatial section ${\bf K}_k$. Unlike the EdS map studied so far, this implies that the conformal transformation to the E-frame:
\eqn\cofw{
ds^2_{k=0, -1} = \Omega(x)^2 \,ds^2_{{\rm E}_d}\;,
}
has singularities even on the $t=0$ spatial section, at those points on ${\bf S}^{d-1}$ where the infinite boundary of ${\bf K}_k$ is mapped.  In particular, a maximally symmetric condensate on ${\bf R} \times {\bf K}_k$ of the form
\eqn\codm{
\bra \phi \ket_k \propto M^{d-2 \over 2}
}
is mapped to an E-frame field 
\eqn\esin{
\bra \tphi(x) \ket_k \propto \left(\Omega(x) M \right)^{d-2 \over 2} \;,
} 
with nontrivial space-time profile, and sharing  the  singularities of the Weyl function $\Omega(x)$. 
The  configuration \esin\  solves the E-frame effective equations of motion with a relevant perturbation
\eqn\srcc{
{\widetilde V}_\Delta [\tphi] = - \left(M \,\Omega(x)\right)^{d-\Delta} \,\left(\tphi(x) \right)^{\, {2\Delta \over d-2}}\;.}
 The physical interpretation in the E-frame is that of an inhomogeneous injection of energy with sharp divergences at the singularities of the conformal map. This happens for $k=0$ at a single point on ${\bf S}^{d-1}$, whereas the infinite injection of energy occurs in the $k=-1$ case along the complete equatorial ${\bf S}^{d-2}$ which separates ${\bf S}^{d-1}$ into two hemispheres. It follows that  singularities of `firewall' type are expected in the global bulk description of such states, in agreement with the philosophy  expressed in \refs\mvrr. The price we pay for the ability to use a classical set up is the need to realize the state in a CFT perturbed by a large relevant operator, but the take-away message ends up being the same.

The behavior of homogeneous condensate states described in this section should admit a natural extension for small perturbations around these states, such as finite-particle excitations. On the other hand, it would be
interesting to generalize the present purely classical description to the full quantum theory. The presence of strongly time-dependent couplings makes the problem challenging. Fortunately,  a number of structural properties of the complementarity maps can be studied in a simplified quantum mechanical model, where time-dependent couplings can be studied at considerable ease.

\newsec{Conformal Quantum Mechanics}

\noindent

In order exhibit these ideas in an explicit quantum framework we can study the quantum mechanical version of
the Landau--Ginzburg models associated to conformal complementarity maps.  A natural construction arises as the $d\rightarrow 1$ limit of the above, in which we replace the classical $d$-dimensional  conformal dynamics of the LG collective degree of freedom  by its $d=1$ analog. It turns out that this simple procedure is somewhat non-trivial, since the different frames  will be found to retain some characteristic features in the $d=1$ system.  

The basic building block is given by the de Alfaro--Fubini--Furlan (AFF) Conformal Quantum Mechanics (CQM) with
Hamiltonian \refs\aff\
\eqn\daff{
 H(\pi, \phi)_{\rm AFF} = \half\left(\pi^2 + {\lambda \over \phi^2}\right)\;,
 }
 for one LG-type degree of freedom $\phi$ with canonical momentum $\pi$.  
  The conformal group acts on the Hilbert space of this theory as an $SL($2,{\bf R}$)$  algebra generated by the Hamiltonian \daff, the dilatation operator $D= \shalf \{\phi, \pi\}$ and the special-conformal generator $C= \shalf\phi^2$, with commutation relations
 $$
 [D, H] = 2iH\;, \qquad [D,C]=-2iC\;,\qquad [H,C] = -iD\;,
 $$ 
 which follow from the basic canonical Heisenberg algebra $[\phi, \pi] = i$. 
 
 The AFF Hamiltonian is classically bounded-below for  repulsive potentials with $\lambda>0$. Even when the potential 
becomes attractive, it remains well defined at the quantum level as long as $\lambda > -1/4$. The spectrum is still well defined for $\lambda > -1/4$ when the system is quantized on $L^2 ({\bf R}^+)$, i.e. on wave-functions $\Psi[\phi]$ with inner product
 $$
 \bra \Psi | \Phi \ket = \int_0^\infty d\phi\,\Psi^* [\phi] \,\Phi[\phi]\;
 $$
 and vanishing boundary condition at the origin, $\lim_{\,\phi\to 0} \Psi[\phi] =0$. 
 More specifically, the Hamiltonian has a positive-definite continuous spectrum \foot{In the absence of a scale, no bound states form.} with delta-function normalization for  
 $-1/4 < \lambda $.
 
  A discrete spectrum can be obtained by placing the system on a   `harmonic trap', i.e. by adding a harmonic potential term with some frequency $\omega$, 
 \eqn\trap{
 H_\omega = H_{\rm AFF} + \omega^2\,C\;,
 }
 where $C = \shalf \phi^2$ is the generator of special conformal transformations. The main advantage of this IR regularization is the preservation of a  nice $SL(2,{\bf R})$ action  on the spectrum, since the Hamiltonian is still linear in the $SL(2,{\bf R})$ generators.  This leads in particular to an equally-spaced discrete spectrum for the trapped Hamiltonian $H_\omega$. 
 
 The trapped models  are analogous to the higher-dimensional conformal field theories defined on spheres, with a gapped spectrum, i.e. the model referred above as the E-frame CFT.  
 The analogy can be sharpened by doing  `dimensional reduction', namely taking the $d\rightarrow 1$ limit of \modlg. The conformal mass term does survive
this limit. The curvature's vanishing is compensated by  the behavior of the conformal coupling $\xi_d$, the product leading to a finite result. Explicitly, one finds for the LG model on ${\bf X}_k = {\bf R} \times {\bf K}_k$
\eqn\trapdr{
\omega_k^2 = \lim_{d\to 1}\, \xi_d \,{\CR}_{{\bf X}_k} = \lim_{d\to 1} \, {d-2 \over 4(d-1)} \,k\,(d-1)(d-2)= {k\over 4} \;
,}
and for the LG model on dS$_d$: 
\eqn\dslimdr{
\lim_{d\to 1}\, \xi_d \,{\CR}_{{\rm dS}_d} = \lim_{d\to 1} \,{d-2 \over 4(d-1)} \,d(d-1) = -{1\over 4}\;.
}
It is interesting that we get the same  {\it tachyonic} `anti-trapping' frequency for the  $d\rightarrow 1$ limits of the hyperbolic and dS theories. This result is natural given the interpretation of the LG models as world-volume descriptions of codimension-one branes on AdS, since we have seen in section 2 that hyperbolic and `bubble' patches of AdS are identical for $d=1$. 

The complete LG action can be derived following the logic  of \refs\kallosh. We can drop a particle probe of mass $m$ in AdS$_{1+1}$ and analyze its near-boundary, slow-motion dynamics in each of the relevant patches:
$$
ds^2_{(k)} = -(r_k^2 + k)\,dt_k^2 + {dr_k^2 \over r_k^2 + k}\;,
$$
in the notation of \manyfol. The particle action reads
$$
S_{(k)} = -m\int dt_k \sqrt{r_k^2 + k -{1 \over r_k^2 + k} \left({dr_k \over dt_k}\right)^2 }\;,
$$
and takes the form of a CQM system with parameters $\omega_k^2 = k/4$ and $\lambda = 2m^2$:
\eqn\cqmk{
S[\phi_k] = \half \int dt_k \left[\left({d\phi_k \over dt_k}\right)^2 - \omega_k^2 \,\phi_k^2 - {\lambda \over \phi_k^2} \right]
}
	in the limit $r_k \gg 1$ and $|dr_k /dt_k  | \ll 1$, where we have used the field redefinition \foot{The canonical mass dimension of the field variable is $-1/2$ in $d=1$, so that the UV/IR relation between $\phi$  and the AdS radius is inverted,  the asymptotic AdS region corresponding now  to the small-field regime.}
\eqn\fruno{
\phi(t_k) = \left({4m \over r_k (t_k)}\right)^{1/2}\;.
}

Although the probe-brane derivation is very transparent, its logical relation to a well-defined AdS$_2$/CFT$_1$ duality is still far from clear. The asymptotic boundary conditions in AdS$_2$ are very sensitive to back-reaction from any finite-energy perturbation \refs\frag\ and the most likely interpretation of the AdS$_2$/CFT$_1$ correspondence involves a large Hilbert space with exactly degenerate states on the CFT side \refs\sen, whose precise relation to AFF-like models is an open problem. We shall not deal with such subtleties in this paper, our aim being more modest. Namely we use the AFF model as a quantum arena to study the eternal/apocalyptic map, while at the same time offering a tentative bulk interpretation of the results.

 \subsec{Deformations And Bound States}
 
 \noindent
 
 We may thus consider three different versions of the CQM model. The standard AFF model with $\omega^2 =0$ (no trapping) 
will be regarded as the analog of  the M-frame, whereas the model with positive trapping $\omega^2 = 1/4$ corresponds to the E-frame. Finally,  the model
with tachyonic anti-trapping $\omega^2 = -1/4$ will be interpreted as the dS-frame (or equivalently the hyperbolic frame).  
More generally, we can deform the AFF model (either trapped or untrapped) by adding a relevant operator contributing to the potential energy as
\eqn\relef{
V_{\Delta} (\phi) = \varepsilon \,{M^{1-\Delta} \over \phi^{2\Delta}}\;,
}
with $\Delta < 1$ (the trapping harmonic term being the particular case $\Delta = -1$). 

We notice that positive  relevant deformations with $\varepsilon >0$ and $\Delta <0 $ behave qualitatively like the trapping term \trap, in the sense that they remove
all the large-$\phi$  `scattering states' near zero energy. Hence, we interpret the models with such a strongly relevant deformation as completely gapped. 
For $\Delta =-1$ we have the strict harmonic trapping, analogous to the E-frame CFT. For $\Delta <-1$ they present a steeper wall, mimicking  a confining potential with a gap proportional to $M$ as $M\gg 1$. Since the complete large-$\phi$ region is removed from the spectrum,  we suggest to interpret such `confining' models as analogous to a sharp wall where  AdS$_2$ is terminated, as in a `bubble of nothing' \refs{\wbn, \dsconf}. 

On the other hand, relevant deformations in the window $0 < \Delta < 1$ are very mild at large values of $\phi$, preserving the continuum of large-$\phi$ scattering states near zero energy. Therefore, we interpret these deformations as leaving behind a sort of `IR CFT fixed point', such as the effective field theory describing  the IR behavior of a system where spontaneous symmetry
breaking has occurred.  In particular,
for $\varepsilon <0$ and large $M$ we find localized classical ground states at $\bra \phi \ket \sim 1/\sqrt{M}$ which we may identify as `condensate' states (cf. figure 7). Such states are analogous to codimension-one brane states propagating in AdS$_2$. 

The $\Delta =1$ case is the marginal deformation. Interestingly, a negative $\varepsilon =-1$ deformation does not automatically imply a global instability of the model, reminiscent of the CdL solutions discussed in the classical models above. The reason is the improved quantum stability\foot{This quantum stability in the small $\phi$ region is further enhanced in the large $N$ limit of the generalization with $O(N)$ symmetry. A term encoding the $N$-dependence of the measure would stabilize the system at large $N$, even in   the absence of a stabilizing  marginal operator.}  which sets the critical value of the effective coupling at $\lambda_{\rm critical} = -1/4$, a phenomenon analogous to the limited tolerance of tachyons in AdS \refs\BF. 

The AFF model admits exact solutions for the condensate states for the particular case of a $\Delta = 1/2$ deformation, since  the resulting induced potential \relef\ is a standard Coulomb interaction. It follows that a spectrum of bound states can be constructed as the radial Hydrogen wave functions continued to real values of the angular momentum, i.e. as (hypergeometric) solutions of
\eqn\hydr{
\left(-\half {d^2  \over d\phi^2} + {\lambda \over 2\phi^2} - {\sqrt{M} \over \phi}\right) U_n (\phi) = E_n \, U_n (\phi)\,}
with discrete spectrum of energies 
\eqn\enerd{
E_n = -{2M \over (2n+1+\sqrt{1+4\lambda})^2}\;, \qquad n\in {\bf Z}_+\;.}
 
  \bigskip
\centerline{\epsfxsize=0.5\hsize\epsfbox{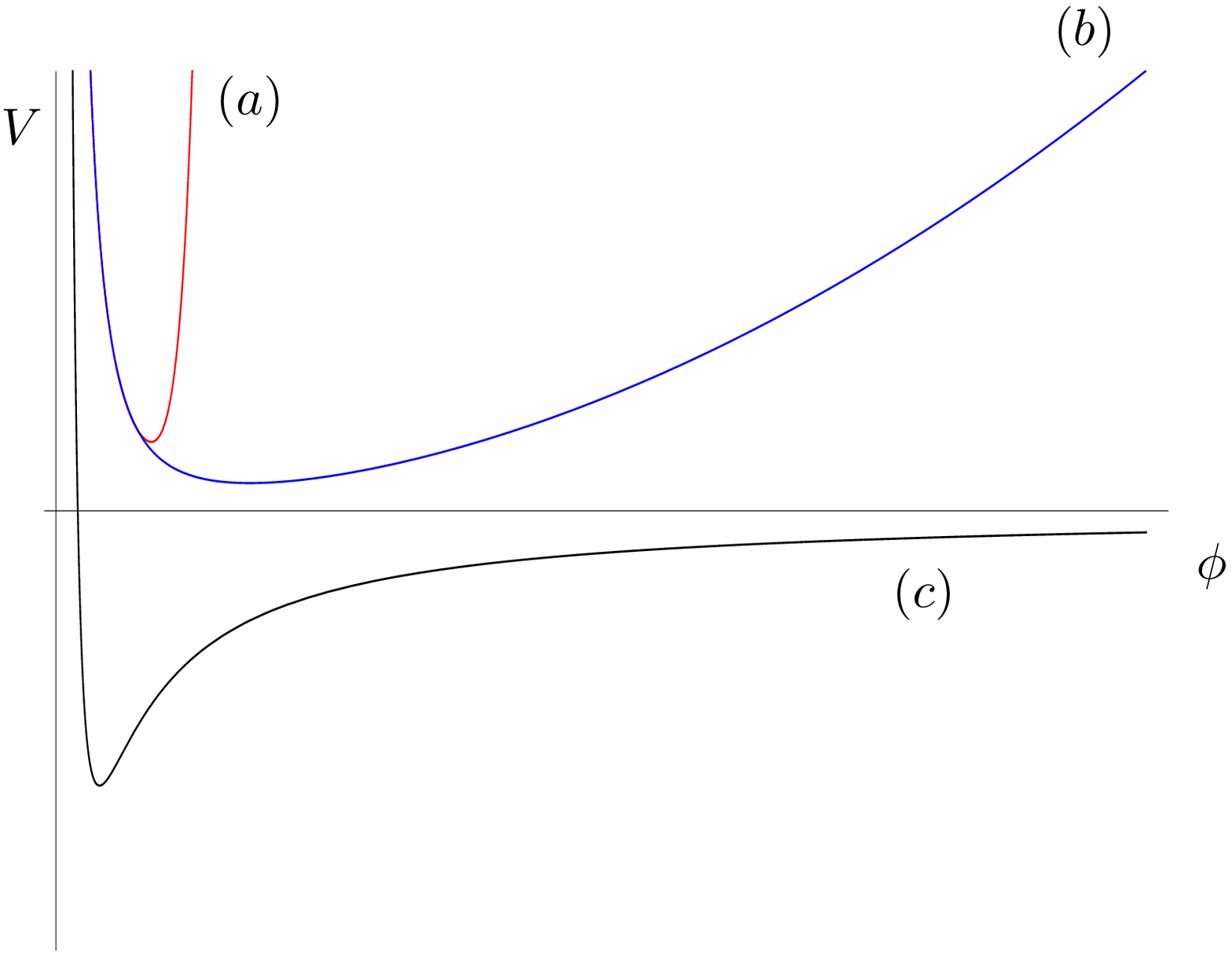}}
\noindent{\ninepoint\sl \baselineskip=8pt {\bf Figure 7:} {\ninerm
The AFF  potential deformed by (a) a positive strongly relevant operator with $\Delta <-1$ (confinement), (b) a harmonic potential, $\Delta=-1$ (trapping),  and $({\rm c}) $
a negative, mildly relevant deformation, $0<\Delta<1$ (condensate). 
}}
\bigskip

The notion of condensate states is inherently semiclassical in the particular case of the dS-frame Hamiltonian, which we write here explicitly, 
$$
H_{\rm dS} = \half\left(\pi^2 + {\lambda \over \phi^2} \right) - {1\over 8} \,\phi^2 - {\sqrt{M} \over \phi}\;,
$$
perturbed by a negative $\Delta=-1$ operator.
The condensate state induced by the last term is 
 necessarily metastable (cf. Figure 8). If this metastable well is deepened  by going to large $M$, the decay probability to the quadratic runaway region is of order $\exp(-a \,M^{2/3})$ for some constant $a$. The bulk interpretation is that of a probe particle which can tunnel out of an accelerating fixed-$\rho$ trajectory, into the low radius region of AdS$_2$. Any such probe that tunnels back to the interior of AdS fails to reach the boundary with infinite E-frame energy, and thus the crunch is prevented. We will return to this intriguing question in section 6. 
 
  \bigskip
\centerline{\epsfxsize=0.5\hsize\epsfbox{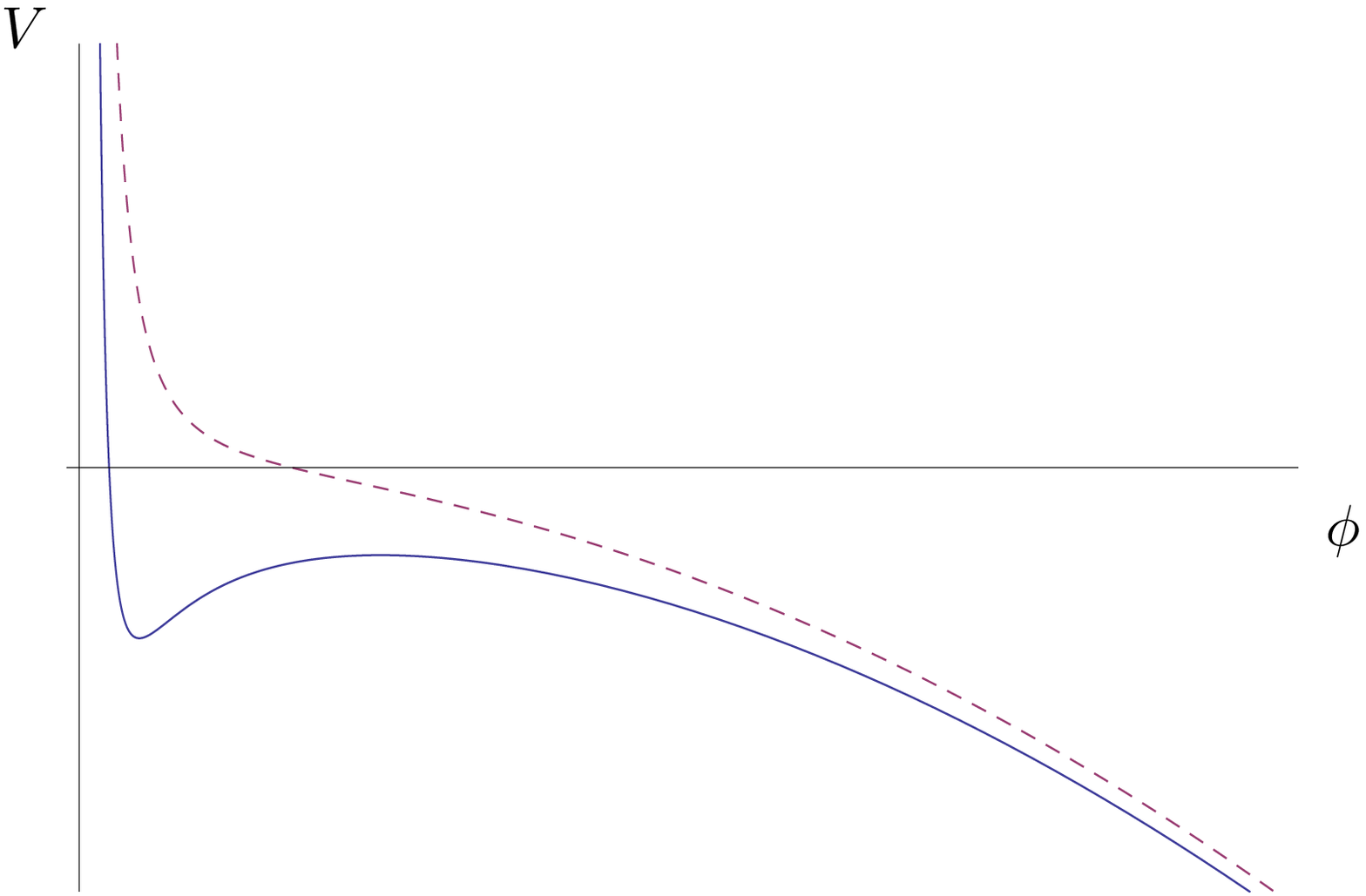}}
\noindent{\ninepoint\sl \baselineskip=8pt {\bf Figure 8:} {\ninerm
The dS-frame potential, with tachyonic anti-trapping (dashed line) and the same model with a relevant operator inducing a metastable condensate (full line).
}}
\bigskip

\newsec{The CQM Complementarity Map}

\noindent

We now study the `conformal complementarity' in the CQM model. For $d=1$, it reduces to the conformal transformation induced by the time-diffeomorphism 
$$
 dt = {d\tau \over \Omega(\tau)}
 $$
 acting as a map between `eternal' time evolution, $\tau \in {\bf R}$, and `apocalyptic' time evolution,  $t\in [-t_\star, t_\star ]$. 
At the classical level we seek the appropriate Weyl function $\Omega(t)$ which maps the E-frame version of CQM:
\eqn\apofa{
{\widetilde S} = \int dt  \left[\half \left({d\tphi \over dt}\right)^2 - {1\over 2} \tomega^{\,2} \tphi^{\,2} - {\lambda \over 2\tphi^{\,2}}\right]\;,
}
with $\tomega^{2} = 1/4$, into the two canonical models of `eternal' type:
\eqn\etfa{
S= \int d\tau \left[ \half \left({d\phi \over d\tau}\right)^2 - \half \omega^2 \,\phi^2 - {\lambda \over 2\phi^2}\right]\;,  }
where $\omega^2 =0$ for the M-frame CQM and $\omega^2 =-1/4 $ for the dS-frame CQM (we use the same time variable for both eternal models for simplicity of notation). 

The answer is obtained by direct substitution of the conformal rescaling $\phi(\tau) = \tphi(t) \, \sqrt{\Omega(t)}$ into the actions. We find the required behavior up to a boundary term: 
\eqn\btermsdu{
S = {\widetilde S} -\int d\tau {dK\over d\tau} = {\widetilde S} - \int dt {d{\widetilde K} \over dt}\;,
}
where\foot{In a slight abuse of notation, we shall often denote the functions $K(\tau) = {\widetilde K} (t)$ with the same letter and drop the twiddle  in ${\widetilde K} (t)$, just as we do with the Weyl factor $\Omega(\tau)={\widetilde \Omega}(t) \rightarrow \Omega(t)$.  } 
\eqn\bterms{
K(\tau) = -{1\over 4} \,\phi^2 \,\pt_\tau \log \Omega(\tau)\;, \qquad {\widetilde K}(t) = -{1\over 4} \, \tphi^{\,2} \,\pt_t \log \Omega(t)\;,
}
provided the Weyl function  satisfies 
\eqn\trasfr{
\tomega^{\,2} = \Omega^2 \, \omega^2 + \half \Omega \,\pt^2_\tau \Omega - {1\over 4} (\pt_\tau \Omega)^2\;,
}
a relation that we may interpret as an `anomalous' transformation law for the frequencies. It is useful to define 
\eqn\ano{
\CA\equiv \half \left( \Omega\, \omega^2 - \Omega^{-1} \,\tomega^{\,2} \right) 
}
for future use, as a measure of such anomalous scaling behavior. In this notation, \trasfr\ reads
\eqn\anos{
\CA = {1\over 8} \Omega^{-1} \left( \pt_\tau \Omega\right)^2 - {1\over 4} \pt_\tau^2 \Omega = {1\over 8} \Omega^{-1} \left(\pt_t \log \Omega \right)^2 - {1\over 4} \Omega^{-1} \pt_t^2 \log \Omega\;.
}

Plugging into   \trasfr\  the actual values of the frequencies, we find two solutions of the non-linear differential equation which, not surprisingly, exactly match the time diffeomorphisms \edst\ and \fc\ found in the context of purely geometrical considerations in AdS$_{1+1}$.

 We have the EM map between the E-frame and M-frame systems, i.e. between the trapped and ordinary AFF models:
\eqn\daffp{
{\rm EM}: \;\;\;\omega^2=0\;, \qquad  \tomega^{\,2} = {1\over 4}\;, \qquad \Omega_{\rm EM} = \half(1+\tau^2) = {1\over 2\cos^2 (t/2)}\;.
}
 The second solution is the standard EdS map, between the trapped and tachyonic versions of the AFF model: 
 \eqn\edsp{
{\rm EdS}:\;\;\;\omega^2 =-{1\over 4}\;, \qquad \tomega^{\,2} = {1\over 4}\;, \qquad \Omega_{\rm EdS} = \cosh(\tau) = {1\over \cos(t)}\;.
}
A useful parametrisation of the two Weyl functions at once is
\eqn\wft{
\Omega_\alpha (t) = {1\over \alpha} \left({1 \over \alpha \cos (t/\alpha)}\right)^\alpha
\;,}
where $\alpha =1$ for the EdS map and $\alpha=2$ for the EM map.

Notice that the singularities of $\Omega(t)$ occur at $t=\pm t_\star$ with $t_\star = \alpha \pi/2$, in agreement with the geometrical features of AdS$_{1+1}$ Penrose diagrams, showing that the Minkowski patch covers a larger portion of the AdS boundary as compared to the dS (hyperbolic) patch (cf. Figure 4).

A relevant operator deformation of the form 
\eqn\relo{
\int d\tau \,{M^{1-\Delta} \over  \phi^{2\Delta}} 
}
in the eternal frame transforms into an analogous term
\eqn\reloa{
\int dt \, {{\widetilde M}^{\;1-\Delta} \over \tphi^{\,2\Delta}}
}
in the apocalyptic frame, where the mass parameters are related by
\eqn\mast{
 {\widetilde M}  = \Omega\,M\;.
 }
 Notice that the map between \relo\ and \reloa\ works also for time-dependent mass parameters, and \mast\ implies that either
 $M$ or ${\widetilde M}$ must be time-dependent in one of two frames.

\subsec{Quantum  Map}

\noindent

The field redefinition between the eternal and apocalyptic frames is generalized to a full quantum map by a correspondence between wave functions
$$
\Psi[\phi, \tau] \longrightarrow \tPsi[\tphi, t]
$$
given by the explicit transformations
\eqn\canouno{
\tPsi[\,\tphi, t\,] = \Omega(t)^{1 \over 4} \,e^{i{\widetilde K}(t) } \,\Psi\left[ \tphi\sqrt{\Omega(t)}, \tau(t)\right]\;,
}
and its inverse 
\eqn\canodos{
\Psi[\,\phi, \tau\,] = \Omega(\tau)^{-{1 \over 4}} \,e^{-iK(\tau)} \, \tPsi\left[\phi/\sqrt{\Omega(\tau)}, t(\tau)\right]\;.
}
In these expressions, $\tau(t)$ and its inverse give the appropriate time diffeomorphism transforming the eternal and apocalyptic frames. The first factor in \canouno\ and \canodos\ is a Jacobian accounting for the correct normalization of both wave functions and the phase is the result of the boundary term in time \btermsdu. It can be checked explicitly that this  map sends solutions of the apocalyptic Schr\"odinger equation 
\eqn\schapo{
i\pt_t \tPsi[\tphi, t] = {\widetilde H}\left(\tphi, -i{\pt \over \pt\tphi}\right) \tPsi[\tphi, t]\;,
}
into solutions of  the eternal Schr\"odinger equation
\eqn\schete{
i\pt_\tau \Psi[\phi, \tau] = H\left(\phi, -i {\pt \over \pt \phi}\right)\,\Psi[\phi, \tau]\;,
}
and viceversa, where the two dual Hamiltonians are defined as 
\eqn\twohals{
H= \half \pi^2 + \half \omega^2 \,\phi^2  + V(\phi)\;, \qquad {\widetilde H} = \half {\widetilde\pi}^{\,2}  + \half \tomega^{\,2}\,\tphi^{\,2}  + {\widetilde V}(\tphi\,)\;.
}
with 
\eqn\defpis{
\pi= -i{\partial \over \pt \phi}\;, \qquad  {\widetilde \pi} = -i{\partial \over \pt \tphi}\;.
}

 The quantum map \canouno\ and \canodos\ is formally a canonical transformation 
 $$
 (\phi, \pi) \longrightarrow (\tphi, \tpi) = \left({1\over \sqrt{\Omega}} \;\tphi,  \sqrt{\Omega} \;\tpi\right)\;,
 $$
 a change of variables that holds in quantum averages up to some anomalous terms coming form the boundary terms. 
We have the basic rules
\eqn\genmapo{
e^{iK} \;\pi \;e^{-i K} = \Omega^{-{1 \over 2}} \left( {\widetilde \pi} + \shalf  \tphi \,\pt_t \log \Omega\right)\;, \qquad e^{-i \widetilde{K}} \;{\widetilde \pi} \;e^{i\widetilde{K}}= \Omega^{1 \over 2} \left(\pi -\shalf \phi\,\pt_\tau \log \Omega\right)
}
which allow us to formulate the general map of observables for general polynomial functions of the canonical operators. In average values defined by 
\eqn\averdef{\eqalign{
\left\langle F(\phi\;;\;{ \pi}\,)\right\rangle_{\Psi} &\equiv \int_0^\infty d\phi \,\Psi^*[\phi, \tau] \,F(\phi, -i\pt_\phi) \,\Psi[\phi, \tau] \;, \cr
\left\langle F(\tphi\;;\; {\widetilde \pi}\,)\right\rangle_{\tPsi} &\equiv \int_0^\infty d\tphi\,\tPsi^* [\tphi, t]\,F\left(\tphi, -i\pt_{\,\tphi}\right)\,\tPsi[\tphi, t]\;.
}}
we have 
\eqn\opemap{\eqalign{
\left\langle F(\tphi\;;\; {\widetilde \pi}\,)\right\rangle_{\tPsi} &= \left\langle F\left(\Omega^{-{1 \over 2}} \phi\;;\; \Omega^{1 \over 2} (\pi -\shalf  \phi \,\pt_\tau \log \Omega) \right)\right\rangle_{\Psi}
\cr
\left\langle F(\phi\;;\;{ \pi}\,)\right\rangle_{\Psi} &= \left\langle F\left(\Omega^{1 \over 2} \tphi\;; \;\Omega^{-{1 \over 2}} ({\widetilde \pi} +\shalf \tphi\, \pt_t \log \Omega) \right)\right\rangle_{\tPsi}
\;.}}
These two equations can be used to extract information about the behavior of any observable. 

A consequence of the quantum map defined above is the complementarity of time evolutions in the respective `eternal' and `apocalyptic' frames. In particular, we can explicitly check the non-commutativity of  the time evolution operators,
\eqn\teo{
{\widetilde U}_t= T_t \, \exp\left(-i\int_0^t dt'\,{\widetilde H(t')}\right)\;, \qquad U_\tau = T_\tau \,\exp\left(-i\int_0^\tau d\tau' \,H(\tau') \right)\;,
} 
 by directly showing that the respective Hamiltonians fail to commute, even at the initial surface where $\Omega=1$. We can compute $[H, {\widetilde H}]$ by expressing, say ${\widetilde H}$ in  eternal-frame variables as 
$$
{\widetilde H} = \half\,{\widetilde \pi}^{\,2} + \half \,\tomega^{\,2}\,\tphi^{\,2}  +{\widetilde V}(\,\tphi\,)= \half\Omega^{-1} \, \tomega^2 \,\phi^2 + \Omega \left(\half \pi^2+ V(\phi)\right)\;.
$$ 
Hence we have the identity
$$
\Omega^{-1} {\widetilde H}= H + \half \left( \Omega^{-2} \tomega^2 - \omega^2 \right) \phi^2\;,
$$ 
from which we find  the commutator of the Hamiltonian operators 
\eqn\comh{
\left[H, {\widetilde H}\right] = 2i\,\CA\,D = 2i\,\CA\, {\widetilde D}\;.
}
in terms of the function $\CA$ defined in \ano\ and \anos, and the dilation operator 
\eqn\dilo{
D= \shalf \{ \phi, \pi\} = \shalf \{\tphi, \tpi \} = {\widetilde D}\;.
}
Since $\CA \neq 0$ even for $\Omega=1$, the two Hamiltonians do not commute in general. Note that they {\it do} commute on those states which are annihilated by the generator of the scale transformations. 
 This interpretation makes contact with the fact that the vacuum AdS manifold realizes the boundary conformal group as an isometry group, showing that the bulk geometry is actually codifying the quantum state of the dual CFT. 

It is interesting to notice that the scale-invariant wave function, satisfying $D\Psi_0 =0$, is given by
$
\Psi_0 (\phi) =  \phi^{-1/2}
$
and is {\it not} normalizable, i.e. it does not sit in the Hilbert space of bound states. Still,  its norm  is less divergent than that of a plane wave. 

\subsec{States And Observables}

\noindent

The quantum map \canouno\ and its inverse \canodos\ are completely general, valid for any state with arbitrary wave-function.  Both versions of the complementarity map have a Weyl function which smoothly tends to the identity  near the origin of times, $\Omega =1$ for $t=\tau=0$, and blows up at the `ends of time',  with a pole-like behavior  $\Omega(t) \sim (t-t_\star)^{-\alpha}$ near $t=\pm t_\star = \pm \alpha\pi/2$.
 
In order to further fix the intuition about the meanings of the quantum map, we can consider a 
 smooth $\tau$-static wave function in the  eternal quantum mechanics, with width $\Gamma$ and centered around $\phi_0$. Its dual to the apocalyptic frame  has a narrowing width ${\widetilde \Gamma}(t) = \Gamma/\sqrt{\Omega(t)}$ as $t\rightarrow \pm \alpha\pi/2$, with its center migrating to the origin as $\tphi_0 (t) = \phi_0 /\sqrt{\Omega(t)}$, while at the same time the phase oscillates wildly. Therefore, the $\tPsi$ wave function is infinitely squeezed into the UV region (small $\tphi$)  as we approach the `apocalypse'. 

Conversely, starting with  $t$-static wave function with fixed width ${\widetilde \Gamma}$  and centered at $\tphi_0$  in the E-frame system, it corresponds to  an eternal wave function slipping into the deep IR (large $\phi$),  trailing the peak at $\phi_0 (\tau) = \tphi_0 \sqrt{\Omega(\tau)}$,  and widening at a rate of order $\Gamma(\tau) ={\widetilde \Gamma} \sqrt{\Omega(\tau)}$.

 To be more precise, let us focus on the operator map \opemap\ for the particular cases of interest.  
We shall adopt a terminology rooted in the behavior of the state in the apocalyptic frame (E-frame) as diagnosed by
the average values of polynomials in the canonical operators or natural observables such as the kinetic, potential and total energy. States with smooth E-frame behavior at $t=\pm t_\star$ can be continued beyond the `apocalyptic' times $\pm t_\star$ and will be termed {\it smooth} ($S$), while those with divergent matrix elements will be denoted as {\it singular}. Among the singular states, we shall refer to {\it crunches} ($C$) when the E-frame potential energy plummets to minus infinity:
$$
\lim_{|t| \to t_\star} \left\bra {\widetilde V} (\tphi\,) \right\ket_{\widetilde C} (t) = -\infty
\;.$$
Singular states with the opposite-sign divergence 
$$
\lim_{|t| \to t_\star} \left\bra {\widetilde V} (\tphi\,) \right\ket_{\widetilde B} (t) = +\infty
\;.$$
will be called {\it bubbles} ($B$) as analogs of `bubbles of nothing'.

The most interesting among singular states are those that look  {\it stationary} in the eternal frame, i.e. with a finite  $|\tau| \rightarrow \infty$ limit of $\bra V(\phi)\ket$, such as the `condensate states' considered in section 5. Any such state with a non-zero value of the eternal potential energy has an apocalyptic potential energy diverging  as $\Omega(t) \sim (t-t_\star)^{-\alpha}$, the hallmark of a singular state.

The anomalous transformation terms do affect the scaling of the kinetic energy:
\eqn\kna{
\left\bra \shalf \tpi^{\,2} \right\ket_{\tPsi} = \Omega(t)\left\bra \shalf \pi^2 \right\ket_{\Psi} -{1\over 4}   \pt_t \log\Omega \,\left\bra \{\phi, \pi\}\right\ket_{\Psi} + {1\over 8} \Omega^{-1} \left(\pt_t \log \Omega\right)^2 \left\bra \phi^2 \right\ket_{\Psi}
\;.}
For a condensate-type state in the eternal frame, the three terms in this equation scale as $(t-t_\star)^{-\alpha}$, $(t-t_\star)^{-1}$ and $(t-t_\star)^{\alpha-2}$ respectively. For either the EM or the EdS map, there is always a singular term for generic values of the eternal frame averages, confirming that the eternally stationary state is a singular state in the apocalyptic frame.    The anomalous terms (second and third on the right hand side of  \kna) are subdominant for the EM model ($\alpha=2$), but have the same scaling  as the first term in the EdS case ($\alpha=1$). \foot{This opens up the possibility that a certain dS-eternal state could be tuned to have finite kinetic energy in the E-frame, although other observables will still diverge in general.}

Starting with a smooth state in the E-frame, with finite and generic values of apocalyptic observables at $t=t_\star$,  the corresponding large-time behavior in the eternal frame follows from the inverse transformations. The 
potential energy vanishes as $\bra V(\phi)\ket \sim \Omega^{-1} \rightarrow 0$  as $|\tau| \rightarrow \infty$ for any value of $\alpha$ (recall this potential energy excludes the purely quadratic trapping term, to be considered below). We say that the state {\it dilutes} away in the eternal frame. 

The inverse of the \kna\ relation 
\eqn\knali{
\left\bra \shalf \pi^{\,2} \right\ket_{\Psi} = \Omega^{-1} \left\bra \shalf \tpi^2 \right\ket_{\tPsi} + {1\over 4}\Omega^{-1} \pt_t \log\Omega \,\left\bra \{\tphi, \tpi\}\right\ket_{\tPsi} + {1\over 8} \Omega^{-1} \left(\pt_t \log \Omega\right)^2 \left\bra \tphi^2 \right\ket_{\tPsi}
\;,}
implies a similar diluting behavior for the ordinary scaling term of the kinetic energy. The anomalous terms depending on derivatives of $\Omega$ have a potentially interesting behavior, since they scale as $(t-t_\star)^{\alpha-1}$ and $(t-t_\star)^{\alpha-2}$ respectively. Hence, for $\alpha=1$ (EdS map) we do get a divergent contribution to the kinetic energy in the $|\tau|\rightarrow \infty$ limit. 
We can understand this behavior by recalling that the EdS map relates the trapped CQM in the E-frame to
the anti-trapped (i.e. tachyonic) CQM in the dS-frame. This is characteristic of the $d=1$ case and implies
that smooth states look as falling down a harmonic cliff in the eternal frame. The potential energy coming from the trapping also diverges as
$$
-{1\over 8} \Omega(\tau) \,\bra \tphi^2 \ket_{\widetilde S} \longrightarrow -\infty\;,
$$
canceling the kinetic-energy infinity coming from the last term in \knali. Hence, the total energy does stay finite in the eternal runaway state. 

Even for $\alpha=2$, i.e. the EM map, the anomalous terms produce an interesting behavior, since the last one
yields an asymptotically constant value of the potential energy. This is to be understood as a state running away to large values of $\phi$, with asymptotically constant kinetic energy, i.e. a standard scattering state in the AFF quantum mechanical model. 

We conclude that the quantum complementarity map sends smooth E-frame states (which do not look `apocalyptic' in this frame) into states which run away towards large $\phi$ values in the eternal frame, with finite total energy
but diverging kinetic and potential components in the particular case of the EdS map.

Starting with a stationary state in the eternal frame, modeled as a `condensate' in the terminology of section 5,
the apocalyptic description carried by the $\tphi, \tpi$ operator algebra, sees it as a singular state. We say it is a crunch when the divergent potential energy is negative, and a bubble of nothing when it diverges to positive infinity.  
\subsec{Evanescent Crunches?}

\noindent

As previously indicated, the EdS map has peculiar properties related to the tachyonic character of the dS-frame Hamiltonian. 
Recall that the EdS map sends the standard trapped CQM (E-frame system) with Hamiltonian 
\eqn\et{
{\widetilde H} = \half \left(\tpi^2 + {\lambda \over \tphi^2} \right) + {1\over 8} \tphi^2 \;,
}
into the tachyonic CQM (dS-frame system) with Hamiltonian
\eqn\ea{
H = \half\left(\pi^2 + {\lambda \over \phi^2} \right) - {1\over 8} \phi^2 
}

The large-$\phi$ instability of the eternal-frame CQM implies that the standard negative deformations of the dS Hamiltonian 
$$
V_\Delta (\phi) = -{M^{1-\Delta} \over \phi^{2\Delta}}
$$
  with constant $M$ and $0<\Delta <1$, fail to induce a absolutely stable condensate at $\bra \phi \ket \sim M^{-1/2}$ with large $M$. In fact, the resulting states are only metastable, with a decay width of order $\Gamma \sim M \exp(-aM^{2/3})$ for some $\CO(1)$ constant $a$. Since the eternality of the condensate is related to the {\it crunchy} character of the state in the apocalyptic frame, it is interesting to inquire whether this metastability, inducing a finite life-time for the condensate, is capable of regularizing the crunch singularity. 
  
We can approximate the very-large $\tau$ wave-function of such states as
\eqn\vlt{
\Psi_{\rm meta} \approx e^{- \Gamma \tau/2} \Psi_{\rm cond} + \sqrt{1-e^{-\Gamma\tau}} \;\Psi_{\rm run}
}
where $\Psi_{\rm cond}$ is a normalized state which solves the Schr\"odinger equation in the large $M$ limit and represents the condensate
in the absence of the tachyonic instability, and $\Psi_{\rm run}$ is a state representing the runaway down the inverted harmonic potential 
after tunneling through the barrier. We can define this running state as the eternal dual from some generic finite-energy state in the E-frame system. 

Upon transforming this wave function to the E-frame system, the $\tPsi_{\rm cond}$ component has the characteristic crunchy behavior we mentioned above, whereas $\tPsi_{\rm run}$ is a smooth state in the apocalyptic frame. The amplitude of the crunchy component does vanish
in the $t\rightarrow \pi/2$ limit as
$$
e^{- \Gamma \tau/2} \sim |\,t_\star - t\,|^{\Gamma /2}
\;.$$
Since we have seen characteristic observables to diverge as inverse powers of $t-t_\star$, the contribution of $\tPsi_{\rm cond}$ to
expectation  values is of order
\eqn\exep{
|\,t-t_\star\,|^{\Gamma - b}
\;,}
where $b$ is some positive constant of  $\CO(1)$ whose detailed value depends on the particular observable being evaluated. In the semiclassical limit where this description of the tunneling is valid, 
 we have $\Gamma \sim M \exp(-a\,M^{2/3}) \ll 1$, so that $\Gamma \ll b$ and the quantum depletion of the condensate is not fast enough to turn off the crunchy behavior of the state. On the other hand, if the depletion rate should become of $\CO(1)$, the exponent
in 
\exep\ could change sign and the corresponding expectation value be smoothed out. We see that the potential for a quantum-mechanical smoothing of the crunch exists, by considering `condensates' with sufficiently fast decay rate. Formally, this situation can be engineered by tuning $M \ll 1$ in units of the background curvature. While the notion of `condensate' is not well defined in such a limit, it is worth mentioning that such states do exist in the dual AdS description (cf. the model discussed in the appendix of \refs\mald) and they have the same qualitative behavior as the more obvious crunch states described here. \foot{In fact, their bulk description is even simpler, since they can be analyzed as weak perturbations of AdS in supergravity.}

\newsec{Generalized Duality Between Eternity and Apocalypse}

\noindent

The description of conformal complementarity maps as quantum canonical transformations in CQM can be 
 {\it formally} extended to higher-dimensional field theories. Consider two conformally related $d$-dimensional Riemannian manifolds ${\bf X}$ and ${\bf\tX}$ with Weyl rescaling function $\Omega(x)$. Let us define a LG model on ${\bf X}$ with classical action 
\eqn\clac{
S_{\bf X}= -\int_{\bf X} \left[\shalf |\pt \phi |^2 +\shalf \,\xi_d \,\CR_{\bf X} \,\phi^2 +V( \phi)\right]\;.
}
The relevant perturbations
\eqn\mscal{
V(\phi) = \sum_i \varepsilon_i\,M_i^{\,d-\Delta_i}\,\phi^{{2\Delta_i \over d-2}}\;.
}
depend on  mass scales $M_i$. This model can be rewritten as a perturbed LG model   on ${\bf \tX}$ with action 
\eqn\clacd{
{ \widetilde S}_{\btX}= -\int_{\btX} \left[\shalf |\pt \tphi |^2 + \shalf\,\xi_d\, \CR_{\btX} \,\tphi^2 +
{\widetilde V} (\tphi\,) \right]\;,
}
plus some boundary terms. The
new potential reads  
$$
{\widetilde V} (\tphi\,) = \sum_i \varepsilon_i\,{\widetilde M}_i^{\,d-\Delta_i} \,\tphi^{\,{2\Delta_i \over d-2}}\;,
$$
in terms of  rescaled point-dependent mass scales ${\widetilde M}_i = \Omega \,M_i$ which now become `source-terms', and with the basic field redefinition $\tphi = \Omega^{d-2 \over 2} \,\phi$. 

The boundary terms are defined as 
$$
S_{\bf X}= { \widetilde S}_{\btX} -\Delta{\widetilde K} = {\widetilde S}_{\btX} +{d-2 \over 4} \int_{\pt \btX} \epsilon \ \,\left(\Omega^{-1} \, {\widetilde \nabla} \,\Omega\right)\,\tphi^{\,2}\;,
$$
where $\epsilon = -1$ for a space-like boundary component and $\epsilon = +1$ for a time-like boundary component and ${\widetilde \nabla}$ is the covariant derivative on ${\btX}$.

For the particular case of a Weyl rescaling function which is only dependent on time,  and a compact spatial section ${\bf K}$,  we can regard the ${\bf X}$ manifold as a cosmology 
$$
ds^2_{\bf X} = -d\tau^2 + \Omega(\tau)^2 \;d\ell^2_{\bf K}
\;,$$
while ${\btX}$ is a static cylinder with base ${\bf K}$:
$$
ds^2_{
\btX} = -dt^2 + d\ell^2_{\bf K}\;.
$$
If $\Omega$ maps a finite interval $t\in [-t_\star, t_\star]$ into the real line $\tau\in {\bf R}$ we can use the terminology that has become standard along this paper and regard ${\bf X}$ as the `eternal frame' and ${\btX}$ as the `apocalyptic frame'. Then, we only have
  space-like boundaries at $t=\pm t_\star$,
so that $\Delta {\widetilde K} = {\widetilde K} (t_\star) - {\widetilde K}(-t_\star)$, with 
\eqn\bttt{
{\widetilde K}(t)  = {d-2 \over 4}  \,\pt_t \log\,\Omega \int_{\bf K} \tphi^{\,2}\;.
}
The boundary term \bttt\ plays no significant role at the classical level, but does feature in the quantum treatment. The formal construction of the states and the canonical map parallels the previous formalism explained in section 5.2, except that Schr\"odinger-picture  wave-functionals replace wave-functions and canonical momenta are defined in terms of functional derivatives in the usual formal fashion, 
$
\pi(x) = -i \delta / \delta \phi(x)
$ and $\tpi(x) = -i\delta/\delta \tphi(x)$. This leads to analogous quantum maps generalizing \opemap:
\eqn\opemapd{\eqalign{
\left\langle F\left[\tphi\;;\; {\widetilde \pi}\,\right]\right\rangle_{\tPsi} &= \left\langle F\left[\Omega^{{d-2 \over 2}} \phi\;;\; \Omega^{2-d \over 2} \left(\pi +{d-2 \over 2}  \phi \,\Omega^{d-1}\pt_\tau \log \Omega\right) \right]\right\rangle_{\Psi}
\cr
\left\langle F\left[\phi\;;\;{ \pi}\,\right]\right\rangle_{\Psi} &= \left\langle F\left[\Omega^{2-d \over 2} \tphi\;; \;\Omega^{{d-2 \over 2}} \left({\widetilde \pi} -{d-2 \over 2} \tphi\, \pt_t \log \Omega\right) \right]\right\rangle_{\tPsi}
\;,}}
with an entirely similar interpretation as their $d=1$ counterparts. 

The Hamiltonian complementarity presented in \comh\ generalizes as well. The eternal-frame Hamiltonian can be written in the form
\eqn\efha{
H = \half \int_{\bf K} \Omega^{1-d}\,\pi^2  + \int_{\bf K} \Omega^{d-1} \left[\half \Omega^{-2} |\pt \phi |^2_{\bf K} + \half \xi_d \,\CR_{\bf X} \,\phi^2 + V(\phi) \right]\;,
}
and its apocalyptic counterpart:
\eqn\apha{
{\widetilde H} = \half \int_{\bf K} \tpi + \int_{\bf K} \left[ \half |\pt \tphi |^2_{\bf K} + \half \xi_d \,\CR_{ \btX} \,\tphi^{\,2}  + {\widetilde V} (\tphi\,)\right]
\;.}
Explicit calculation then shows that 
\eqn\cham{
\left[ H, {\widetilde H} \right] = 2i \,\CA(t)\, {\widetilde D}\;,
}
where ${\widetilde D} = \shalf \int_{\bf K} \{\tphi, \tpi\}$ and the anomalous  term generalizes to \foot{It is interesting that $\CA$ vanishes for $d=2$, a case that we have excluded consistently from our discussion in this paper. Precisely at $d=2$ we expect genuinely {\it anomalous} (quantum) contributions to \cham.}
\eqn\ant{
\CA(t) = {d-2 \over 4} \pt_t^2 \log \Omega + {(d-2)^2 \over 8} \left( \pt_t \log \Omega\right)^2 \;.
}
Although these relations are derived by simple canonical manipulations, we expect them to hold in all generality for general field theories, showing that the Hamiltonian complementarity induced by conformal mappings trivializes when acting on scale-invariant states, annihilated by the dilation operator. It is natural to interpret this result as underlying the fact that the global AdS vacuum is invariant under the action of the dilation operator, represented in the bulk as an isometry. On the other hand, for any state with an intrinsic scale, we expect a non-trivial quantum complementarity between the two Hamiltonian evolutions.

\newsec{Discussion}

\noindent

In this paper we have analyzed aspects of the general idea, going back at least to \refs{\banksdo, \insightfull}, that
complementarity maps can be realized as conformal transformations (or more general field-redefinitions) in
holographic models. A particularly simple example of this program was proposed in \refs\usd, in terms of condensate states on perturbed CFTs defined on dS space-time. A conformal map to the same CFT on the Einstein universe (E$_d$), but now perturbed by a time-dependent coupling, serves as the `infalling' frame in the sense of horizon complementarity. 

We have singled out the change of time variables, from an eternal history in dS$_d$, to a finite or `apocalyptic' one in E$_d$, as an `UV remnant' of the complementarity map, which can be studied using conventional Lagrangian methods. We have done so at the level of classical Landau--Ginzburg models of brane-like states, extending the analysis already presented in \refs\usd. A full quantum analysis is possible for the $d=1$ version, conformal quantum mechanics, which retains some qualitative properties akin to a AdS$_2$ dual, despite the fact that no actual reconstruction of bulk dynamics is available. In this case, the quantum map is a canonical transformation which rescales the canonical operator basis by a time-dependent factor. Once identified, the construction can be formally extended to higher dimensions. 

The existence of two operator algebras: $\{\CO\}$  for the  `exterior' observables and  $\{{\widetilde \CO}\}$ for  the `infalling' observables,  is similar to the case
of electric/magnetic duality or T-duality in string theory. In order to sharpen the analogy, we must enrich the models by allowing the coexistence of both `windings' and `momenta'. So far we have considered extremely simple states by way of example, such as either dS condensates or E-frame stationaries.
In order to realize the quantum complementarity map in a more physical fashion we must introduce an appropriate `measuring apparatus' for each operator algebra. 

Consider for instance a dS condensate state. Any physical system constructed from stationary states around the condensate ground state will only measure the eternal properties of the dS state. On the other hand, a physical system whose physical size is `comoving' with the Hubble expansion, will be able to `measure' a crunch in the $t$ time variable. It is tempting to us  to use our own universe as an example (in the limit $G_{\rm N} \rightarrow 0$ with fixed Hubble constant).
The Higgs condensate has fixed size in units of the Hubble constant, but a comoving `observer', anchored on the `realm of the nebulae',  will measure the 
$\{{\widetilde \CO}\}$ operator algebra. At large $\tau$-times such an observer is necessarily made of `neurons' separated by super-horizon distances, so that its workings appear completely non-local to an observer furnished with the $\{\CO\}$ operator algebra.  Conversely, in its own frame the $\{{\widetilde \CO}\}$ observer will see the
$\{\CO\}$ observer as a shrinking entity whose own Hamiltonian ramps up the eigenfrequencies to  produce the illusion of eternity in the face of an impending crunch.

The main limitation of these considerations is the absence of an actual reconstruction of operators with
approximate bulk locality, in the spirit of \refs{\banksdo, \bena, \lifsh, \polch,\papado}. In this sense, we have
strived to characterize horizon complementarity in the absence of an actual `horizon', using only deep UV data. The dichotomy between local and strongly non-local observables in the CFT should then become even more drastic when translated to reconstructed bulk operators.

\bigskip{\bf Acknowledgements:} 

The work J.L.F. Barbon  was partially supported by MEC and FEDER under a grants FPA2009-07908 and FPA2012-32828, the Spanish
Consolider-Ingenio 2010 Programme CPAN (CSD2007-00042), Comunidad Aut\'onoma de Madrid under grant HEPHACOS S2009/ESP-1473 and the 
spanish MINECO {\it Centro de Excelencia Severo Ochoa Program} under grant SEV-2012-0249. 
The work of E. Rabinovici  is partially supported by the
American-Israeli Bi-National Science Foundation,  the Israel Science Foundation Center
of Excellence and the I Core Program of the Planning and Budgeting Committee and The Israel 
Science Foundation "The Quantum Universe".

{\ninerm{
\listrefs
}}

\end